\documentstyle[12pt,aaspp4,flushrt]{article}
\def\etal{{et~al.}~}
\lefthead{Lewis \etal}
\righthead{X-ray Mass Estimates at z$\sim$0.3 for the CNOC Cluster Sample}
\begin{document}

\title{X-ray Mass Estimates at z$\sim$0.3 for the CNOC Cluster Sample}
\author{Aaron D. Lewis\altaffilmark{1} and E. Ellingson\altaffilmark{2}}
\begin{center}
\affil{Center for Astrophysics and Space Astronomy, \\ University of Colorado
at Boulder, Campus Box 389, Boulder, CO, 80309}
\altaffiltext{1}{lewisad@colorado.edu}
\altaffiltext{2}{e.elling@casa.colorado.edu}
\author{Simon L. Morris\altaffilmark{3}}
\affil{National Research Council of Canada, Herzberg Institute of
Astrophysics, Dominion Astrophysical Observatory, 5071 West Saanich Road,
Victoria, B.C., V8X 4M6, Canada}
\altaffiltext{3}{Simon.Morris@hia.nrc.ca}
\and
\author{R. G. Carlberg\altaffilmark{4}}
\affil{Department of Astronomy, University of Toronto, Toronto, Ontario, M5S
1A7 Canada}
\altaffiltext{4}{carlberg@moonray.astro.utoronto.ca}
\end{center}

\begin{abstract}

Results are presented from the analysis of ROSAT HRI and PSPC observations of the CNOC
subsample of the EMSS high redshift galaxy clusters. X-ray surface brightness profiles of 14
clusters with $0.17 < z < 0.55$ are constructed and fit to isothermal $\beta$ models.  Where
possible, we use both the HRI and PSPC data to constrain the fit.  Under the assumptions of
isothermality, hydrostatic equilibrium, and spherical symmetry, we derive total X-ray masses
within a range of radii from 141 to $526 h^{-1}_{100}$~kpc.  These masses are compared with
both the dynamical masses obtained from galaxy velocities and the projected masses from
published gravitational lensing studies. We find no systematic bias between X-ray and
dynamical methods across the sample, with an average 
$M_{Dyn}/M_X=1.04\pm0.07$, although individual clusters exhibit mass discrepancies up to a
factor of 2. We estimate that the systematic effects due to cooling flows, non-equilibrium
systems and temperature gradients affect the average mass ratio by no more than
$15-20\%$. Weak gravitational lensing masses appear to be systematically higher than X-ray
results by factors of
$\sim50\%$, while strong lensing estimates show larger discrepancies (factors of
$\sim2.5$).  However, these comparisons are complicated by the need to extrapolate the X-ray
data to larger or smaller radii. X-ray derived cluster gas masses are calculated, from which
we obtain a cluster baryon fraction of
$\sim5\%h^{-3/2}_{100}$, yielding $\Omega_0 \sim0.3h^{-1/2}_{100}$.
\end{abstract}


\section{Introduction \label{secint}}
Clusters of galaxies are the largest gravitationally bound systems in the universe, and hence
serve as critical regions for cosmological studies.  This has led to their extensive use as a
probe of the evolution of large scale structure (see \cite{hen97} for a review).   Of
particular interest to cosmologists, clusters may directly constrain the universal density
parameter,
$\Omega$.  Under the assumption that the co-moving volume from which the cluster formed
contains the same fraction of baryons and dark matter as the field, cluster mass-to-light
ratios or cluster baryon fractions yield a value of $\Omega$ (the latter method also requires
a value of $\Omega_{\rm b}$,\cite{whi93}; Carlberg \etal 1996, C96
hereafter; \cite{sad97}).  

Additionally, measurements that constrain the evolution of the cluster luminosity function
and mass function are basic to our understanding of large scale structure in the universe
(c.f. \cite{ouk97}), as well as providing the observational data necessary to constrain
theoretical models of structure formation (\cite{hen97}). The accurate determination of both
total gravitational cluster masses and the mass of baryonic matter in clusters is thus
crucial to cosmology.

Intermediate redshift ($0.2\lesssim z\lesssim 0.6$) clusters are optimal for the study of
cluster properties. They are ubiquitous enough to be found in statistically useful numbers,
they are compact enough to be observed in a single field of view with modern instrumentation,
and they are bright enough to permit detailed investigation. The hot ($10^8$ K), diffuse
($10^{-3}$ cm$^{-3}$) intracluster medium (ICM) may be the most accurate tracer of the
potential well of the cluster.   Following the assumptions of hydrostatic equilibrium and
isothermality, the surface brightness distribution of a resolved X-ray image yields the ICM
density as well as the cluster mass (\cite{cav76}; \cite{bah77}).   The  longevity of the
ROSAT and ASCA satellites has allowed us to compile long pointed exposures of a sample of
clusters at intermediate redshift,  to obtain a statistical comparison of mass determination
methods.

Additional methods for cluster mass determinations include the optical measurement of galaxy
redshifts with subsequent virial/dynamical analysis,  and the measurement of both weak and
strong gravitational lensing of background galaxies by the cluster mass potential.
Theoretically, all three methods should accurately provide the same results, yet there have
been discrepancies reported among the techniques (e.g. \cite{car94}; \cite{fah94};
\cite{mir95}; Wu \& Fang 1997, \cite{wu97} hereafter; Allen 1998, A98 hereafter; Wu \etal 1998,
\cite{wu98} hereafter).  Each method is subject to possible systematic errors. Our goal is to
analyze a well-defined sample of clusters and identify any discrepancies between the three
methods, including any systematic biases. Therefore we have performed an X-ray analysis of a
sample of 14 clusters which have been observed optically and have a  detailed
dynamical/virial analysis. In addition, many of the clusters have gravitational lens
measurements published for comparison.

In \S \ref{secdat} we describe the cluster sample and the data.  We present the data
reduction methods in \S \ref{secredux}. Analysis of the data is in \S
\ref{secanal}, with discussion in \S \ref{secdis}, and conclusions in \S
\ref{secconc}. The Appendix contains notes on individual clusters and their analysis.

This paper assumes $H_0 = 100 h$~km~s$^{-1}$~Mpc$^{-1}$ and $q_0 = 0.1$ unless otherwise
noted.

\section{Cluster Sample \& Data \label{secdat}}
We have chosen to analyze the available X-ray
data on clusters within the CNOC (Canadian Network for Observational Cosmology, Yee,
Ellingson, \& Carlberg 1996) intermediate redshift subsample of 15 EMSS (Extended
Medium Sensitivity Survey) distant clusters (\cite{gio90b}; \cite{hen92}).  The CNOC sample
also includes the optically selected cluster Abell 2390, another rich, X-ray luminous
cluster.   Spectroscopic observations were performed at the CFHT 3.6~m telescope at Mauna
Kea by the CNOC collaboration, who obtained 30-200 galaxy velocities per cluster, as well as
accurate photometry.  For a discussion of the selection criterion for this sample, see
\cite{yee96}. The dynamical analyses of these data are presented in \cite{car96} and
\cite{car97}(1997).

Here we present an analysis of the  available ROSAT X-ray observations of this sample. To
resolve the X-ray core structures of clusters at these redshifts, the
$\sim5\arcsec$ FWHM resolution of the High Resolution Imager (HRI)  is necessary. At a
redshift of 0.3,
$5\arcsec~\simeq~15h^{-1}$~kpc.  Thus the $\sim30\arcsec$ FWHM of the ROSAT Position
Sensitive Proportional Counter (PSPC) cannot resolve structure smaller than
$\sim90h^{-1}$~kpc, making its use problematic with clusters with small core radii.

The HRI is a mature instrument, and all 16 of the CNOC clusters have been the subject of
pointed observations by us and other investigators.   Two of the CNOC clusters, 
MS~1231.3+1542 and MS~1621.5+2640, have HRI observations which do not allow us to
extract adequate surface brightness profiles, and are not discussed in this work. We present
here results from the 14 remaining clusters in the CNOC sample.  These HRI images comprise
the primary data sets for our analysis.

In addition, we have analyzed ROSAT PSPC images of 5 of the CNOC clusters from the public
archives.  Additional serendipitous PSPC observations of MS~1231.3+1542 and MS~1512.4+3647
are available, but are of insufficient duration for our analysis. A combined analysis of the
two data sets, where available, is utilized to yield a more accurate model of the X-ray
surface brightness profiles. Table 1 lists the clusters in the sample with their
identification, redshift,  and L$_x$ in the {\it Einstein} IPC band, in columns (1), (2), and
(3), respectively.  Columns (4) and (5) list the total duration of the HRI and PSPC
observations used in our analysis, respectively.
\placetable{tableone.tex}

\section {Data Reduction \label{secredux}}
\subsection {HRI data sets}

We use the IRAF/PROS\footnote{The Image Reduction and Analysis Facility (IRAF) is
distributed by the National Optical Astronomy Observatories, which is operated by the
Association of Universities for Research in Astronomy, Inc. (AURA), under contract to the
National Science Foundation.  The Post Reduction Off-line Software (PROS) package was
developed by the High Energy Astrophysics Division of the Smithsonian Astrophysical
Observatory, with assistance from the Space Science Computing Division and the Laboratory for
High Energy Astrophysics of the Goddard Space Flight Center.}  
, STSDAS\footnote{The Space Telescope Science Data Analysis Software Package (STSDAS)
software is distributed by the Space Telescope Science Institute (STScI) software development
group.  STScI is operated by AURA
for the National Aeronautics and Space Administration.}
, and ESAS\footnote{\raggedright{The Extended Source Analysis Software (ESAS) package created
by Steve Snowden (\cite{sno95}) is available at the ROSAT GOF,
ftp://legacy.gsfc.nasa.gov/rosat/software/fortran/sxrb/}}
analysis packages to reduce the data. Each observation is inspected for periods of
unacceptably high count rates, which are primarily due to cosmic ray events or pointing of
the spacecraft too close to the Earth or Sun.  Standard pipeline analysis by the ROSAT data
facility succeeds in eliminating nearly all such events.  The majority of each data set
consists of
$0.5-2$~ksec blocks of continuous pointed observation.  We eliminate any block under 50~s to
avoid periods of unstable or unreliable pointing and peculiar timing effects. After time
filtering, multiple observations of the same cluster are combined, and the resulting data set
is checked to ensure accurate mapping of sky coordinates, as some observations are performed
months or years apart.

The HRI has a pixel scale of $0.5\arcsec$~pixel$^{-1}$, and the instrument efficiency within
a radius of $5\arcmin$ varies by less than 10\%
\footnote{\raggedright{Details may be found in ``The ROSAT High Resolution Imager (HRI)
Calibration Report'', available at
http://hea-www.harvard.edu/rosat/rsdc\_www/HTI\_CAL\_REPORT/hri.html}}
(partly owing to the intentional `wobble' of the spacecraft during each observation). For all
of our clusters, the detected cluster emission does not extend beyond this region.  However,
to include a large enough area to allow background measurements, we perform our analysis over
a region 700 pixels in radius. Thus we employ the ESAS software to create exposure and
particle background maps that take into account time, energy and pixel-to-pixel variation in
the detector during the observation.  Images of each cluster (with X-ray surface brightness
contours overlaid to emphasize structure) are presented in Figure
\ref{figpp}. The X-ray center of each cluster is identified  by iteratively fitting a two
dimensional 5~pixel FWHM gaussian to a blocked (by a factor of 8), smoothed (with a gaussian
of 1~pixel FWHM), particle-background and exposure corrected image to find the highest peak. 
Cluster centers generally coincide within $1\arcsec$ of the brightest pixel on the X-ray
image,  and within $5\arcsec$ of the brightest cluster galaxy (BCG) identified optically
(c.f. Gioia \& Luppino 1994, see the Appendix for a discussion of individual offsets). Point
sources are similarly identified, and their positions and extent are recorded in an image
mask, which is later used to avoid contamination of the surface brightness profile, or the
flux calculation (see \S \ref{secanal}).
\placefigure{figpp}

\subsection{PSPC data sets}

The PSPC has time, energy, and pixel dependent variations of the instrument efficiency and
the detector resolution which are much greater than those affecting the HRI.  Accurate
exposure, background, and scattered solar X-ray maps in each energy band must be used to
properly recover the signal in an image. We have used the ESAS package to create images which
are corrected for these effects.   We restrict our analysis to the PSPC bands R4 through R7,
also known as SASS band B, corresponding to an observed energy band of $0.44-2.04$ keV. We
identify cluster centers and any point sources in a similar fashion to the HRI reduction. 

\section{Data Analysis \label{secanal}}

\subsection{Surface Brightness Profiles}

After the initial reduction described in the previous section, each image is binned into
radial annuli about the cluster center (165 annuli for HRI data, 50 for PSPC) out to a
maximum radius (700~pixels for HRI, 50 for PSPC), and a radial surface brightness profile is
constructed by summing the counts in each bin.

In some cases, the X-ray emission is detectably elliptical or irregular. In the former case a
more acceptable fit may be found using elliptical, rather than radial, surface brightness
profiles.  However, the use of elliptical coordinates is problematic, in that the
position angle of elliptical isophotes changes with increasing radius introducing an
additional free parameter to the fit, which can be difficult to constrain with
statistical significance for our noisier datasets.  As discussed in \cite{neu97} and
B\"{o}hringer \etal (1998), the difference in the derived mass caused by an ellipticity
such as that evident in Abell 2390 is less than 5 percent compared with that derived from a
radially symmetric profile.  We note that in the case of MS~0015.9+1609, which exhibits
significant non-circularity,  a radial and elliptical analysis has been performed by
\cite{neu97}, and the fitted $\beta$ parameters, and hence the masses, for the two models are
identical. We have chosen to use radial profiles throughout this work for
simplicity, consistency in analysis between clusters, and for our comparison with the
dynamical masses (which also assume sphericity). Other uncertainties in the analysis (e.g.
in the temperature) dominate any error introduced by assuming radial symmetry. 

The projected radial surface brightness profile is then fit  via a non-linear least squares
routine to a standard isothermal
$\beta$-model (\cite{cav76}; \cite{bah77})
\begin{equation}
\label{eqbeta}
S(b) = S_0[1+(b/r_{core})^2]^{-3\beta + 1/2} + C,
\end{equation}
where $S_0$ is the central surface brightness, $r_{core}$ is the cluster core radius, $\beta$
is the ratio of energy per unit mass in the galaxies to that in the gas, and $C$ is the
background.  Note that we fit a uniform background assumed to be cosmic in origin, which is
in addition to the particle background we have derived with the ESAS software and previously
subtracted.

As has been noted by several authors (e.g \cite{jon84}), an isothermal
$\beta$ model is inaccurate for clusters out of equilibrium (e.g. undergoing a merger or
sub-merger), or those with large temperature gradients.   Clusters in our sample in general
exhibit regular X-ray and optical morphology and galaxy velocity distributions, although a
few of the clusters do show signs of recent merging activity. We discuss these and other 
possible effects on our results in \S \ref{secdis}. 
 
Based on either narrow-band optical or X-ray images, several clusters have been identified or
are candidates for large central cooling flows  (c.f. Donahue, Stocke, \& Gioia 1992, D92
hereafter; Pierre \etal 1996; \cite{all96}; Rizza \etal 1997), in some cases accounting for
25\% of the total observed X-ray luminosity.  In an X-ray image, a cooling flow may be
expected to be seen as a sharp peak in emission concentrated in the core of a cluster, where
the cooling time for the ICM is shorter than the current age of the universe at the cluster
redshift.  

One problem introduced by a cooling flow is its effect on our attempt to fit a
$\beta$ model to the observed surface brightness profile. If a cooling flow is present in the
cluster one would expect to see an uncharacteristically high central peak in the radial
profile, or a second shoulder at smaller radius than the shoulder related to the
$r_{core}$ parameter. If the shoulder of the cooling flow is indistinguishable from the
underlying profile, it will result in a lower fitted value of the $r_{core}$ parameter. 
Unfortunately, uncertainties in the fit to the $r_{core}$ parameter are directly correlated
with those in the $\beta$ parameter, and a low fitted value of
$r_{core}$ will result in a low fitted $\beta$ value.

To partially avoid the problems of lower fitted $r_{core}$ and hence lower fitted
$\beta$ values, we  have attempted to eliminate the effect of any peaked cooling flow
emission from the fit.  In appropriate cases, we have used a method similar to the pioneering
work of Jones \& Forman (1984) who excluded central regions of clusters with excess emission,
in order to obtain an acceptable fit.  We subjectively identified an inner cutoff radius
based on a significant increase in emission at a radius smaller than the obvious shoulder of
the profile, and excluded data within that radius from the least-squares fit to the $\beta$
model.  In some cases, the cutoff radius is not obvious, although a cooling flow may still be
present. In Figure \ref{figsb} we present the HRI surface brightness profiles, overlaid with
best fit $\beta$ models when all points are included  in the fit (solid line) and only points
outside the chosen cutoff radius are included (dotted line). The inner cutoff radius, where
applicable, is indicated in Figure \ref{figsb} with a vertical dashed-dotted line.
\placefigure{figsb}

Observations at other wavelengths may also imply a cooling flow, for instance H${\alpha}$
filaments (\cite{hec89}), or emission line signatures in the spectra of the brightest cluster
galaxy (BCG). Based on the observed [\ion{O}{2}] emission from the CNOC observations
(\cite{bal97}) and a study of extended H${\alpha}$ in EMSS clusters by \cite{don92} 
we have attempted to identify the existence and estimate the strength of a possible cooling
flow in each cluster.  We found that [\ion{O}{2}] emission  in our BCG spectra is not well
correlated with apparent cooling flow size.  However, all large cooling flows in the sample
had significant [\ion{O}{2}] emission, and non-cooling flow clusters did not.  Thus, the
presence of [\ion{O}{2}] emission in the galaxy spectra can  be regarded as an indicator of
possible cooling, though not a reliable  measure of the strength of the cooling flow. 

\cite{don92} detected extended H${\alpha}$ emission on scales of tens of kpc for
several clusters in our sample. In Figure \ref{figha} we have plotted the ratio of
H$\alpha$/X-ray  luminosity versus the inner cutoff radius.  We choose the ratio of
luminosities to account for the fact that a cooling flow of a given H${\alpha}$ luminosity
will have a greater perturbation  on the profile of a poor cluster than on a rich one.  There
appears to be a mild correlation between  this ratio and the empirical inner cutoff radius
described above.   However, we do not feel this correlation should be used to estimate the
cooling radius in lieu of inspection of a surface brightness profile; rather we regard the
correlation as an independent justification for our identification of
which clusters have significant cooling flows, and our subjective  choice of cutoff radius. 
\placefigure{figha}

High spatial resolution is important to constrain both 
$r_{core}$ (especially at these redshifts), and the effects of cooling flows, while low
background counts are needed for a more accurate measurement of the faint slope in the tail
of the surface brightness distribution, which constrains $\beta$. As discussed in
\cite{riz97}, HRI observations may not allow an accurate measurement of the slope of the tail
of the cluster profile, due to the high background of the instrument.  The slope of the
surface brightness profile is the primary constraint on the fit to the $\beta$ parameter, and
a high and irregular background may mask out the signal of an increasing slope, resulting in
a fit to a flattened slope and correspondingly low fitted
$\beta$ value. Even if we have minimized
the effect of a cooling flow on the $\beta$-model fits, we are still faced with the high
background of the HRI. To improve our fits, we have utilized PSPC data
where available. The PSPC data are in general not sufficient to resolve the cores of the
clusters.  However, they do provide a lower background count rate, allowing cluster surface
brightness profiles to be accurately measured to greater radii than with HRI data alone. 
We attempted use a combined profile to more accurately describe the true shape of the cluster
emission than would be possible with a single instrument.  First we constrain the $r_{core}$
parameter with the HRI data, and then use it as a fixed value when fitting to the PSPC data,
from which we derive final $\beta$ values. In cases where we excluded inner regions from the
HRI fit to avoid cooling emission, we also excluded the same region from the subsequent PSPC
fit.  Parameter values from the HRI, PSPC, and the combined data sets are shown in Table
\ref{tabletwo.tex},  columns (2) \& (3), (4) \& (5), and (6) \& (7), respectively. In this
way, we hope to most effectively utilize the strengths of each instrument.  We note that the
PSPC fits obtained solely with PSPC data (independent of the HRI $r_{core}$ value) are in
general agreement with the HRI's independent fits (see the Appendix for individual
cases).  
\placetable{tabletwo.tex}

\section{Discussion \label{secdis}}
\subsection{Cluster Mass Determinations}

Under the assumption of an isothermal sphere in hydrostatic equilibrium (\cite{sar88}), where
the gas and the dark matter are identically distributed, we can calculate the total X-ray
derived gravitational mass within a radius $r$ 
\begin{equation}
\label{eqmx}
M_X(<r)=1.13\times10^{15}\beta\biggl[\frac{T}{10\rm{keV}}\biggr]
\biggl[\frac{r}{\rm{Mpc}}\biggr]
\frac{(r/r_{core})^2}{(1+(r/r_{core})^2)}h^{-1}\rm{M}_{\sun},
\end{equation}
where $T$ is the isothermal gas temperature. The parameters $\beta$ and $r_{core}$ were
obtained as described in the previous section.  The cluster temperatures are listed in
Table \ref{tabletwo.tex}, column (8).  The majority have been obtained from pointed ASCA
observations of each cluster analyzed by \cite{mus97}. For MS~0906.5+1110 we use the
L$_x$--T$_x$ relationship from 
\cite{mus97} to estimate the gas temperature because no direct measurement is available. 
The uncertainty in the temperature is estimated from the scatter in the relation.

For MS~0302.7+1658 and MS~1006.0+1202 we use the empirical Bgc-T$_x$ relation. The quantity
Bgc is a statistical measure of the galaxy overdensity within the cluster (c.f.
\cite{yee87}).  This quantity is well-correlated with both the velocity dispersion and T$_x$
for these clusters (see also \cite{ylc99}). Since it is based purely on the photometry of
galaxies in the cluster, it is essentially an independent estimate of the cluster richness;
which, unlike L$_x$, is not affected by cooling flows.
Uncertainty in the temperature values are estimated from the scatter in the
Bgc-T$_x$ relation (\cite{yee99}).

Table \ref{tablethree.tex} lists the inner cutoff radius for each cluster (column 2), the
cluster radius where the HRI signal drops to 2 $\sigma$ above the noise (column 3), and the
total X-ray derived mass within this radius  (column 4).  We have plotted the total mass
profiles for each cluster in Figure \ref{figmp} (solid curves).  Also listed in Table
\ref{tablethree.tex} are the 2 $\sigma$  outer cutoff radii for the PSPC data and the masses
within these radii (columns 6 \& 7, respectively).  Uncertainties in the masses are 1
$\sigma$ statistical errors associated with the
$\beta$ model parameters and the temperatures.
\placetable{tablethree.tex}

Of primary interest is a comparison of total cluster mass estimates from this method with
both dynamical/virial analyses and gravitational lensing techniques.  Many authors have
reported lensing masses in excess of the results of other techniques (e.g.
\cite{fah94}; \cite{mir95}).
\cite{wu97} find good agreement between dynamical and lensing methods for a large sample
including eight of the CNOC clusters; however they report X-ray calculated masses a factor of
2-3 lower than lensing mass estimates.  \cite{all97} performed a detailed analysis
of a nearly identical subsample of the clusters analyzed by \cite{wu97}, including a cooling
flow analysis, concluding that strong lensing and X-ray mass estimates are in good agreement
for strong cooling flow clusters, which they identify as the only truly relaxed systems.

We have included in Table \ref{tablethree.tex} a mass comparison between the X-ray and the
dynamical/virial methods (our comparison with gravitational lensing is discussed in \S
\ref{secgrav}). Virial mass estimates for each cluster are taken from
\cite{car96}, and scaled to our 2 $\sigma$ cutoff radius according to the observed galaxy
density distribution, which 
\cite{car97}(1997) found to be well-fit by  a Navarro profile (\cite{nav96}) of the form:
\begin{equation}
\label{eqpg}
\rho_{gal}(r^{\prime}) = \frac{\rho_0}{(r/r_{200})(0.27 + r/r_{200})^2}
\end{equation}
where $r_{200}$ is the radius at which the cluster density is 200 times the critical density,
as tabulated in Carlberg, Yee, \& Ellingson (1997, C97 hereafter), and $r^{\prime}$ is simply
equal to
$(r/r_{200})$. The total dynamically derived gravitational mass enclosed within a radius $r$
is then given by:
\begin{equation}
\label{eqmo}
M_{Dyn}(<r)=\frac{0.82~h^{-1}M_{200}}{9.56}\int\limits_0^r 4\pi (r^{\prime})^2
\rho(r^{\prime})\,dr^{\prime}
\end{equation}
where the factor of $0.82\pm0.14$ is an empirically derived correction by
\cite{car97ip} to the virial mass estimate based on the more detailed dynamical mass
estimate, and $M_{200}$ is the total mass enclosed at $r_{200}$ (also tabulated in
\cite{car97ip}).   The factor of 9.56 is the normalization of equation \ref{eqmo}
required to yield $M=M_{200}$ at $r=r_{200}$.  Note that our cosmology has been chosen to
match that of \cite{car96}, so that the masses may be directly compared.  
\placefigure{figmp}

We have overlaid the dynamical mass profiles on the X-ray derived mass profiles in Figure
\ref{figmp} (dashed curves).  Note that the basic shapes of the profiles are similar due to
the inherent similarity between a Navarro density profile and an isothermal $\beta$-model
density profile; however the normalization and shoulders of the profiles are constrained by
independently determined parameters.  Across the sample, clusters exhibit a scatter of both
higher and lower X-ray masses relative to the dynamical masses.  In Table
\ref{tablethree.tex}, column (5), we have listed the ratio of dynamically to X-ray derived
mass for each cluster at the HRI 2 $\sigma$ cutoff radii. These mass ratios are also
plotted in Figures \ref{figdmm} and \ref{figdmr} as functions of cluster mass and the 2
$\sigma$ radius, respectively.   Note that we have omitted MS~0906.5+1110 from the
dynamical/X-ray comparison; its spuriously high velocity dispersion invalidates the dynamical
mass calculation (\cite{car96}).   In addition, we have increased the uncertainty associated
with the velocity dispersion for MS~0440.5+0204 to include a recent measurement made by
\cite{gio98}.  
\placefigure{figdmm}
\placefigure{figdmr}

We find that the masses from the dynamical and X-ray methods have a weighted average ratio of
$1.04\pm0.07$ with an rms scatter in the individual points of 0.27.
Statistically, the dynamical/X-ray mass ratios ($M_{Dyn}/M_{X}$ ratios, hereafter) show no
intrinsic scatter above the $\sim 20-30\%$  observational uncertainties.  However, our
pre-selection of rich, X-ray luminous systems, and especially our exclusion of the probable
merging cluster MS~0906.5+1110 (see below), undoubtably plays a role in producing the
relatively small scatter in Figures \ref{figdmm} and \ref{figdmr}. C96 also found a small
scatter in the distribution of dynamical M/L ratios for these clusters, a further indication
of the relative uniformity of the sample. 

We have analyzed the data in Figure \ref{figdmm} to determine if there is a correlation
between the $M_{Dyn}/M_{X}$ ratio and the cluster mass. There appears to be a slight
systematic gradient, with lower masses having slightly lower ratios. We fit the data points
with both a zero slope line and with a free parameter for the slope. The addition of the
slope parameter is only significant at the 1.5~$\sigma$ level and the range of differences
over our sample is about 10\%, similar to the uncertainty in the mean value. We note that the
correlation becomes somewhat weaker if plotted as a function of the radius at which the mass
measurement is made (see Figure \ref{figdmr}). This suggests that if the correlation with
mass is real, it may be due to small systematic differences in cluster properties as a
function of their mass, rather than observational effects. A larger sample will be necessary
to confirm whether this gradient is indeed statistically significant.

\subsection{Systematic Biases in the X-ray Mass Estimates\label{subsecsys}}

The agreement between the independent X-ray and dynamical cluster mass estimates is good,
implying that these cluster mass estimates are unbiased to  within $\sim10$\% systematic
uncertainty. However, a number of possible systematic errors in the X-ray mass estimates must
still be addressed. The effects of the instrumental point spread function (PSF), large-scale
temperature gradients,  cooling flows, and cluster substructure may all contribute
unaccounted-for biases in the X-ray masses.  While the agreement between the two independent
methods is greatly encouraging, the magnitude of these effects must be estimated in order to
place a firm limit on the maximum possible bias in  the cluster  mass estimates. 

\subsubsection{Instrumental PSF}

First we address the effect of the instrumental PSF on the data. For those clusters with the
smallest core radii, the $5\arcsec$ HRI   PSF may increase the fitted $r_{core}$ value (and
the fitted $\beta$ value).  To obtain a first-order estimate of the effect of the PSF, we
simulate cluster data with various core radii and convolve them with a  gaussian of FWHM
comparable to the HRI PSF ($\sim5\arcsec$).  Fitting these data with
$\beta$-models  allows us to estimate the increase in parameters due to the PSF. For the
cluster with the smallest measured core radii (MS~1512.4+3647 with
$r_{core}=9\arcsec$), we estimate a systematic increase of $\sim 46\%$ in
$r_{core}$ and $\sim 6\%$ in $\beta$ due to the HRI PSF.  For the next smallest cluster core
(MS~0302.7+1658 with $r_{core}=10\arcsec$), the estimated increase is $\sim 24\%$ in
$r_{core}$  and $\sim 6\%$ in $\beta$. Continuing to larger core radii, for MS~1455.0+2232 the
effect of the PSF is to increase $r_{core}$ by $\sim 16\%$ and $\beta$ by
$\sim 4\%$. For the remaining clusters, the net effect is no greater than
$\sim 9\%$ for the $r_{core}$ parameter, and $\sim 2\%$ for the $\beta$ parameter.

Since the resolution of the PSPC is comparable to or greater than the core radii in our
sample, we do not use it to measure the core radii, but instead adopt the HRI values for our
iterative fits. We used a similar method as that described above for the HRI to estimate the
effects of the PSPC PSF  on $\beta$. For the MS~1358.4+6245 PSPC data, the final fitted
$\beta$ value is changed by $8\%$ due to broadening by the PSF; all other clusters are
affected by less than 5\%.

Although the HRI PSF can increase the fitted value for
$r_{core}$, and decrease the value of $\beta$, the net effect on the X-ray masses are
actually quite small. This is partly because we calculate the enclosed mass at many
core radii, but also because the effects on the two parameters tend to partially cancel. If
we correct the final fitted values for the cluster sample to account for the PSF of both
instruments, the greatest effect on any of our X-ray derived mass estimates is to decrease it
by $6\%$ (in the case of MS~1512.4+3647). Three other clusters exhibit a decrease of
$3\%$, and the remaining clusters change less than $2\%$.  The overall affect on the weighted
average $M_{Dyn}/M_X$ ratio for the sample is negligible.   Applying corrections for the PSF,
the weighted average ratio for the entire cluster sample is increased by $0.2\%$. On a
subsample of 9 clusters identified as cooling flow clusters (see below), the weighted average
is decreased by $0.4\%$. In both cases this correction is many times smaller than other
possible sources of bias. Thus we choose to present uncorrected values in all tables and
figures. 

\subsubsection{Large-Scale Temperature Gradients}

As mentioned in \S \ref{secredux}, our assumption of isothermality may not always be valid. 
Clusters of galaxies often show evidence for substructure in their X-ray emission and their
distribution of galaxies, both of which could be signatures of non-isothermality. However,
the clusters in this sample include some of the richest and most relaxed systems known. We will
attempt to quantify below the effect that a reasonable temperature gradient might have for
such systems.  We do not expect extreme temperature gradients due to the short equilibration
timescale of the intracluster gas (\cite{dav96}).

Recently, \cite{mar98} used extensive ASCA
observations to construct temperature profiles for 30 nearby ($z\sim0.05$) clusters. They
observed a similarity among all the profiles, with decreasing temperature at larger radii. 
Assuming a uniform temperature gradient comparable to that found for their sample,  we can
estimate the effect of large-scale temperature gradients on our mass estimates.  We
approximate the temperature gradient as a power law across the range of radii at which we
measure cluster masses, and assume the following form:
\begin{equation}
\frac{T(r)}{T_0}=A\biggl(\frac{r}{r_{180}}\biggr)^{-\alpha}
\end{equation}
where $T_0$ is the emission-weighted temperature we have previously assumed, $r_{180}$ is the
cluster radius at which the density equals 180 times the critical density, and $\alpha$ and
$A$ are constants. If a mass density
$\propto r^{-2}$ is assumed, $r_{180}$ equals
$1.054~r_{200}$ . A non-isothermal temperature requires inclusion of a second term in the
X-ray mass calculation that accounts for the temperature gradient.  After replacing the
emission-weighted temperature with the assumed power-law temperature form, we obtain a new
X-ray estimate for the total cluster mass:
\begin{equation}
M_{X_{grad}}(<r)=M_{X_{iso}}A\biggl[\frac{r}{r_{180}}\biggr]^{-\alpha}
\biggl[1+\frac{\alpha(1+(r/r_{core})^2)}{3\beta(r/r_{core})^2}\biggr],
\end{equation}
where $M_{X_{iso}}$ is our original X-ray mass estimate (equation \ref{eqmx}).
From the data in Figure 7 of \cite{mar98}, we obtain values of 0.4 and 0.5 for
$\alpha$ and $A$, respectively. This new mass estimate is higher than the isothermal estimate
at small radii, and lower at larger radii.  The difference from the isothermal mass estimates
is between $-9\%$ and $+29\%$ for individual clusters. For the sample as a whole, the weighted
average $M_{Dyn}/M_X$ ratio is lowered to $0.96\pm0.07$. Therefore, if the clusters in our
sample at moderate redshift exhibit temperature gradients similar to those seen for rich
clusters at low redshift, we can estimate a systematic effect on the average mass
ratio of $\lesssim10\%$ due to large-scale temperature gradients.

\subsubsection{Cooling Flows}

Our X-ray mass estimates of a cooling flow cluster will be biased relative to a non-cooling
flow cluster, as the lower temperature gas in the cooling flow will reduce the value of any
emission weighted temperature measurement that is not spatially resolved (as is the case for
the low spatial-resolution ASCA observations which we rely on).  Multi-phase
temperature models have been employed by some authors to attempt to more precisely determine
the state of the ICM in cooling flow clusters (e.g. \cite{all97};
\cite{boh98}).  However the additional free parameters associated with a multi-phase fit
introduce very high uncertainties in the temperature determinations, and the measured single-phase
temperature value cannot be excluded in most cases.   Spatially resolved high resolution
spectroscopy (as is possible with AXAF, but not yet available) will be needed to adequately
constrain the more complex models.  

We can use the $M_{Dyn}/M_X$ ratios in subsamples of cooling flow and non-cooling flow
clusters to estimate the effects of cooling flows on our results.  The dynamical mass is
independent of any X-ray temperature bias, and thus we expect to see higher ratios for the
cooling flow clusters than for the non-cooling flow clusters due to this effect. We have
divided our sample into cooling and non-cooling flow systems. We have classified all those 
clusters with non-zero inner cutoff radii as listed in Table \ref{tablethree.tex} as cooling
flows, with the exception of MS~1006.0+1202, whose suspected cooling flow is very small. We
see that our cluster sample is dominated with suspected cooling flows, with only five clusters
excepted (including MS~0906.5+1110, which is not included in the averages, and the addition
of MS~1006.0+1202 as a non-cooling flow cluster).
  
Taking weighted averages of the $M_{Dyn}/M_X$ mass ratios found in Table
\ref{tablethree.tex}, we find
$1.00\pm0.08$ and $1.15\pm0.13$ for the cooling flow and non-cooling flow systems,
respectively. Comparing the two sets of ratios with a K-S test shows that there is a 51\%
probability that the two sets come from the same parent distribution.  Thus we do not detect
the expected high ratio for cooling flow clusters, and suggest that this effect is masked by
the random uncertainties in the mass estimates.

In order to investigate the possible effects on the subsample $M_{Dyn}/M_X$ ratios due to
misclassification of cooling flows based on our subjective evaluation of the
azimuthally averaged surface brightness profile, we have repeated our calculation with a more strictly
delineated cluster sample.  We have divided the sample into those clusters with strong detections of
extended H${\alpha}$  ($L_{{\rm H\alpha+[N~II]}}>10^{42}$~ergs~s$^{-1}$, as measured by \cite{don92}),
those with firm non-detections, and those with missing, weak, or indeterminate detections ($L_{{\rm
H\alpha+[N~II]}}<10^{42}$~ergs~s$^{-1}$).
The strong detection sample, which we define as definite cooling flows includes Abell~2390,
MS~0440.5+0204, MS~0839.8+2938, MS~1224.7+2007, MS~1358.4+6245, and MS~1455.0+2232.   
The weighted average $M_{Dyn}/M_X$ ratio for these clusters is $0.98\pm0.09$. We see that
even with a restricted cooling-flow subsample, the mass ratios are consistent with unity, and
we do not infer a large systematic bias due to the emission-weighted temperatures.
The non-detection sample, which we define as definite non-cooling flows includes
MS~0451.5+0250, MS~0906.5+1110, and MS~1008.1-1224.  
The weighted average $M_{Dyn}/M_X$ ratio for these clusters (excluding MS~0906.5+1110) is
$1.17\pm0.19$, which is also consistent with unity, as well as the larger non-cooling flow
subsample identified above.
 
This contrasts with the results of \cite{all97}, who find a large correction to the isothermal
temperature measurement ($\sim 50\%$), and hence the X-ray masses, when using a multiphase
model for the cooling flow clusters in their sample. We note that our cooling flow clusters
may on average be less extreme systems than those investigated by
\cite{all97}, whose cooling flow sample had an average mass deposition rate of $\sim 1700$
M$_{\sun}$~yr$^{-1}$. We also note that our comparison of cluster subsamples does not reveal a
large discrepancy between cooling and non-cooling flow clusters such as that seen by
\cite{all97}, and we estimate that the systematic effect on our average mass ratio from the
temperature measurements of our cooling flow clusters does not exceed $\sim10\%$, similar to
the lowest corrections in
\cite{all97}.

\subsubsection{Non-equilibrium Systems and Cluster Substructure \label{subsecnoneq}}

A possible complication in comparing cooling flow and non-cooling flow clusters comes from
the possibility that all non-cooling flow clusters are significantly  out of equilibrium in
their cores. \cite{all97} suggests that these clusters all have elevated core radii and that
X-ray mass estimates in the core will thus be biased significantly low. The effect of core 
structure should be much smaller on our measurements than for their work, since in general we
measure the masses well outside of the core region. However, we searched for evidence of
systematic differences in fitted parameters and appearance of the images which might indicate
that only the cooling flow clusters are in equilibrium. We do find some differences
between cooling flow and non-cooling flow clusters in our sample. We have labeled those
clusters exhibiting obvious substructure or asymmetry in their HRI images with an asterisk in
Table \ref{tablethree.tex}. Unfortunately, this subjective definition is difficult to apply to
those clusters with very low signal-to-noise ratio images, such as MS~0302.7+1658. Furthermore, we are in
general barely resolving substructure in the cores (being limited by the PSF of the HRI, which is
$\sim15~h^{-1}$~kpc at $z=0.3$), and it is difficult to evaluate the regularity of the core of
each cluster. However, some structure is evident upon visual inspection of the HRI images. Of
the six clusters with the most obvious substructure or asymmetry, only MS~0015.9+1609 is
identified as a cooling flow cluster according to our inner cutoff radius criterion, while
none of the six clusters would be defined as a cooling flow cluster according to our stricter
H$\alpha$ luminosity criterion. Thus clusters in our sample that exhibit irregularity in
their HRI images are almost always identified as indeterminate or non-cooling flow clusters,
consistent with \cite{all97}'s hypothesis.

But to what extent does this irregularity affect our X-ray mass estimates? 
If we divide our sample into clusters with very regular appearance and those with the highest
degree of irregularity, we may compare the $M_{Dyn}/M_X$ ratios to see if there is a
discrepancy.
The most regular clusters are Abell~2390, MS~0440.5+0204, MS~0839.8+2938, MS~1224.7+2007,
MS~1358.4+6245, MS~1455.0+2232, and MS~1512.4+3647.  The irregular clusters are
MS~0015.9+1609, MS~0451.5+0250, MS~0451.6-0305, MS~0906.5+1110, MS~1006.1+1202, and
MS~1008.1-1224. Taking weighted averages of the $M_{Dyn}/M_X$ mass ratios, we find
$0.98\pm0.08$ and $1.18\pm0.12$ for the regular and irregular subsamples, respectively.
Thus the $M_{Dyn}/M_X$ ratios do not appear to be significantly perturbed by substructure or asymmetry.


Nonetheless, non-equilibrium conditions which persist well outside of the cluster core may
still affect our mass estimates.  While the $M_{Dyn}/M_X$ ratios are in unbiased agreement
for our sample, we must ask - could both estimates be wrong due to non-equilibrium? The only
way this may systematically affect our average mass ratio estimates by more than the
$\sim15\%$ difference between cooling and non-cooling flow subsamples is if non-equilibrium
conditions have simultaneously biased both X-ray and dynamical masses by nearly the same
amount. The expectation, however, is that merging and non-spherical systems will affect the
dynamical mass estimates much more than the X-ray estimates, in part because of the longer
relaxation time for the former, but also because of the different nature of the measurements
made. For dynamical mass estimates, a non-spherical and unrelaxed system is  prone to having
an elevated velocity dispersion measurement due to substructure, as well as the increased
difficulty of separating field galaxies from cluster galaxies within a complex velocity
distribution. Confusion from this (or simply from not obtaining sufficient numbers of galaxy
velocities) can easily increase the measured velocity dispersion by factors of 25-100\%, and
the corresponding dynamical masses by factors of 1.5--4. On the other hand, numerical
simulations suggest that X-ray mass estimates will be less affected by substructure in the 
cluster gas.  Extensive numerical simulations by
\cite{evr96} have demonstrated that X-ray observations fitted to isothermal $\beta$ models
produce nearly unbiased cluster mass estimates in most cases. The worst biases, of 15--35\%,
came from measurements at very large radii. Thus we expect that incidences of strong
non-uniformity will not easily  lead to mass ratios near unity, but instead will bias the
$M_{Dyn}/M_X$ ratios high. 

The cluster MS~0906.5+1110 may be considered an illustration of the greater effect of major
mergers on dynamical masses. This cluster was omitted in  our comparison between X-ray and
dynamical masses, as well as the analysis of
\cite{car96}, because its velocity dispersion was double-peaked and implausibly large -- an
excellent candidate for a merging system. Its  $M_{Dyn}/M_X$ ratio was found to be
$2.76\pm1.01$, suggesting that the X-ray mass estimate is much less affected by the cluster's
non-equilibrium state. 

However, it seems that minor mergers may still produce mass ratios close to unity.  The
cluster MS~0451.5+0250 also shows signs of a possible core merger,  with two very bright
BCG-like galaxies and complex X-ray morphology in its center. However, here the galaxy
velocity dispersion does not appear to be exceptionally broad, the X-ray contours are smooth
in the outer regions, and good fits to a
$\beta$ model were obtained from the PSPC data. Its $M_{Dyn}/M_X$ ratio is
$1.23\pm0.29$. This cluster may be undergoing a merger which is affecting only the innermost
parts of the cluster.  In this case it seems more likely that both mass estimates, which are
made well outside of the core regions, are unaffected by  core substructure, rather than that 
they are both strongly biased by nearly exactly the same amount. 

One more set of independent data can be used to check on whether the mass estimates may be
systematically affected by substructure. The total galaxy luminosity should allow a rough
cluster mass estimate which is not strongly affected by  substructure. 
\cite{car96} report that the dynamical M/L ratios for this sample (omitting MS~0906.5+1110)
are consistent with a universal value.  This implies that the random measurement uncertainties
are larger than any gross errors in the individual cluster dynamical masses due to
substructure. We computed the M$_X$/L ratios for our sample at the same radii at which the 
$M_{Dyn}/M_X$ ratios  have been computed, using data from \cite{car96},
and also find the distribution to be consistent with random uncertainties.  In order to
estimate remaining possible systematics in the mass estimates due to substructure, we
compared the average M$_X$/L values for cooling flow and non-cooling flow clusters. If
cooling flow clusters are assumed to be equilibrium systems, then the limit of the difference
in their M/L ratios sets the maximum effect of non-equilibrium on the non-cooling flow
clusters. We found that the ratio of the cooling flow to non-cooling flow M$_X$/L values is
1.19$\pm 0.24$, statistically consistent with no effect, but possibly indicating a small
bias. Thus we estimate the maximum systematic in the non cooling flow clusters due to
non-equilibrium to be $\sim 25\%$.

\subsubsection{Summary of Possible Systematic Biases}

We have found the effect of the instrumental PSF to be negligible on our data set. Following
the assumption of \cite{mar98} that their observed temperature gradient exists in our cluster
sample, we estimate a maximum systematic increase on the weighted average X-ray masses of
$\lesssim10\%$. We have compared cooling and non-cooling flow clusters in our sample and
estimate that a failure to account for the lower emission-weighted temperature in cooling flow
clusters  may decrease the mass ratios for the cooling flow subsample by
$\lesssim10\%$. 

Finally, we have discussed the effects of non-equilibrium conditions, particularly in
non-cooling flow clusters.  If non-equilibrium conditions produce nearly the same bias in
both dynamical and X-ray mass estimates, the excellent agreement between their values may not
indicate that both estimates are fundamentally correct. We argue that this is not likely to
happen in systems with significant merging, as exemplified by the cluster MS~0906.5+1110.
Further, we note that the consistent agreement between dynamical and X-ray mass estimates,
and the total cluster galaxy light makes it difficult to argue that any cluster in our sample
has an estimated mass which is in error by significantly more than the statistical
uncertainties in the measurement. A comparison of cooling flow and non-cooling flow clusters
suggests an upper limit on the average mass bias due to non-equilibrium to be $\sim 25\%$.

Combining these possible systematics allows firm limits to be placed on the possibility that
these cluster mass estimates are significantly biased. For cooling flow clusters, we estimate
a total possible bias  in the X-ray measurements of $-10\%$ to $+10\%$ (for effects of the
cooling flow and large scale temperature gradients, respectively). For non-cooling flow
clusters we estimate the possible biases to be no more than $-25\%$ (for non-equilibrium) to
$+10\%$ (for temperature gradients).  We stress once again that these relatively tight limits
are probably only made possible by the uniformity of the sample and that a sample with
stronger cooling flows, stronger merging systems, or with mass comparisons made closer to the
cluster cores might exhibit larger biases.

We note again that the dynamical masses we compare to have been corrected by $16\%$ from the
virial mass estimates, following \cite{car97ip}. Since this correction is comparable to our
possible systematics, we cannot say unequivocally that it is mandated by this comparison, but
we note that it is fully consistent with our results.

\subsection{Comparison With Gravitational Lensing Measurements \label{secgrav}}

We have also performed a comparison of our masses to the currently available gravitational
lensing masses in the literature.  Using our fitted parameters, we calculate the projected
X-ray mass (based on the derivation in Gioia \etal 1998)
\begin{equation}
\label{eqmxp}
M_{X_{proj}}(<r)=1.74\times10^{15}\beta\biggl[\frac{T}{10\rm{keV}}\biggr]
\biggl[\frac{r}{\rm{Mpc}}\biggr]\frac{(b/r_{core})^2}{\sqrt{1+(b/r_{core})^2}}
h^{-1}\rm{M}_{\sun},
\end{equation}
in a cylinder of the same radius, and in the same cosmology, as the given lensing mass
estimate. The symbols are the same as in equation (2), and $b$ is the projected radius on the
sky. The ratio of gravitational lensing mass to X-ray mass is plotted in Figure
\ref{figgmr} as a function of the lensing radius.   We find that masses from strong arcs are
systematically higher than X-ray masses, by up to a factor of
$\sim$~5, with most values a factor of $\sim2$ higher.  Weak lensing estimates are also
higher, though the discrepancy is less (factors of $\sim1.5$).  
\placefigure{figgmr}

Can we understand which estimators are biased, since they do not agree?  As can be seen in
Figure \ref{figgmr}, many of the strong lensing masses are measured at very small radii, in
some cases near our inner cutoff for cooling flows. In general, these values show the
strongest discrepancies.  The nature of our simple isothermal model implies a smooth mass
distribution into the core of the cluster.  Large mass clumps inside a few tens of kpc will
not be accurately modelled. A98's analysis of cooling flow clusters yielded core radii
typically a factor of 5 smaller than ours (on the order of 25$h^{-1}$ kpc), suggesting  much
more highly concentrated cores than implied by our fits, even after including a correction
for the instrumental PSF. This difference in our fitted parameters bears the primary
responsibility for our differing results, as the X-ray masses are highly dependent on the
value of the core radius when measured at small radii. Clearly the effects of substructure
and asymmetry discussed in Section \ref{subsecnoneq} may affect the X-ray mass estimates more
strongly in the cluster cores, possibly accounting for the discrepancy.

However, we note that at somewhat larger radii the enclosed mass is well-constrained by the
X-ray  profiles, with little dependence on the fitted core radii. We can therefore set strong
upper limits on the total cluster mass at radii greater than about 100$h^{-1}$ kpc.  While a
high lensing  mass may indicate a  strongly peaked mass distribution  in the cluster cores,
(e.g. \cite{mir95}), which the X-ray profile cannot constrain,  a lensing mass within a few
tens of kpc which is greater than the X-ray derived mass at a much larger radius is difficult
to reconcile (e.g. MS~1006.0+1202, see Figure \ref{figgmr}).

One significant source of uncertainty in the strong lensing estimates is the lack of redshift
information available for many of the lensed arcs. Another important difference in mass
estimates may be due to differences in the X-ray and lensing centers of the clusters (see
\cite{all97} for a discussion).  If a large offset (e.g. $> 50$~kpc) exists between the
centers, the two methods will be probing mass distributions through distinctly different
lines of sight through the cluster.  In this case, one may not expect to be able to perform a
meaningful comparison of the cluster mass at small radii. We have listed the sky coordinates
of the X-ray centroids for all of our clusters in the Appendix. For the clusters we have in
common with \cite{lup95}, \cite{pie96}, and \cite{all97}, our cluster centers coincide to
within a few arcseconds. Fahlman \etal (1994), \cite{sma95}, Smail \etal (1997), and
\cite{wu97} do not list the positions they have centered their mass estimates on for
the clusters we have in common with them. Asymmetry or substructure in the lens may
also lead to overestimates of the lensing mass by factors of $\sim1.6$, as shown by
\cite{bar95}. Finally, strong lenses may preferentially be found in clusters which also
have other intervening mass distributions along the line of sight, tending to systematically
increase the lensing mass over the cluster mass.
 
In general we find that the weak lensing estimates are in better agreement with the X-ray and
dynamical masses than are the strong lensing mass estimates, although they  remain
systematically higher. Here the radii of comparison are generally larger than the  2 $\sigma$
radii to which the X-ray mass profiles are measured, and similar to typical radii for
dynamical measurements.  Possible reasons for discrepancies may include  assumptions in the
lensing model about the mass profile in the cluster and uncertainties in the redshift
distributions of the lensed sources.  Our $\beta$-model extrapolation to the quoted
weak lensing radii may also be inconsistent with the assumed mass distribution profiles 
in the different estimates.  We suggest that concurrence may be possible with weak lensing
estimates, and that constraining the remaining systematics in both methods may resolve the
discrepancies. A uniform treatment of the sample is necessary, but unfortunately there is not
yet enough information in the literature to explore the comparisons in such a manner.
Individual comparisons with both strong and weak lensing mass estimates are presented in the
Appendix.

Our results are notably different from those of \cite{wu97}, who compared lensing, virial,
and X-ray mass estimates for a sample of clusters which included seven objects in common with
our sample.   Their analysis assumed fixed values of
$\beta$ and $r_{core}$,  and temperatures primarily derived from the L$_x$--T$_x$ relation.
They concluded that while dynamical masses appeared to be in agreement with both weak and
strong lensing results, X-ray methods underestimated the cluster mass by factors of 2--3.  
However, their chosen sample is quite heterogeneous, and includes several clusters which may
have erroneously high velocity dispersions. This may have biased their dynamical results
toward higher values.

Recently, \cite{wu98} reanalyzed their cluster sample (which now includes nine clusters in
common with our sample), and concluded that the X-ray, weak lensing, and dynamical methods
are all appropriate and accurate tracers of the gravitational potential at many core radii. 
Their dynamical/X-ray mass ratios are $1.08\pm0.7$, in general agreement with our results, but
with much larger scatter. We note that the CNOC dynamical masses appear to have much
less scatter in the $\sigma-T$ relation than other clusters in their sample, and suggest again
that the CNOC sample provides a much more accurate measure of velocity dispersions than is
usually available at these redshifts. We also suggest that the more detailed treatment we
have presented of both dynamical and X-ray mass calculations contributes to the much smaller
scatter we find in the mass ratios.

At smaller radii, \cite{wu98} find large discrepancies between strong lensing and X-ray mass
estimates, similar to our results, and suggest that this is due to failures in the X-ray
analysis to account for the effect of cooling flows on the emission-weighted temperature, or
clusters being out of dynamical equilibrium (especially the cores of non-cooling flow 
clusters), resulting in gross overestimates of the core radius and hence low X-ray mass 
estimates at small radii. We note that for our sample the effects of the cooling flow
temperature bias appear to be at the 10\% level, and we suggest that this is not a viable
explanation for the lensing/X-ray discrepancy in our sample. 
Alternatively, we propose that either systematics in the strong lensing mass
estimates, or non-equilibrium in the X-ray core structures may be the cause of the
discrepancy.

\subsection{Baryon Fractions and $\Omega_0$}

Observations of the ICM X-ray emission also allow us to estimate the gas density and the
total gas mass in each cluster.  Cluster fluxes and luminosities in the ROSAT HRI band
($0.1-2.4~$keV)  are listed in Table \ref{tablefour.tex}, columns (2) \& (3).   We model the
emission as thermal bremsstrahlung according to a Raymond-Smith plasma (\cite{ray77})
consisting of hot gas at the measured cluster temperature, containing 40\% of the
\cite{all73} elemental abundance values, and absorbed by cold Galactic
hydrogen\footnote{$N_H$ values are taken from \cite{dicwww}.  The database can be
accessed via Starview over the world wide web at
http://skyview.gsfc.nasa.gov/cgi-bin/skvadvanced.pl}.
\placetable{tablefour.tex}

The gas density at a radius $r$ was calculated for a $\beta$ model (\cite{jon84}):
\begin{equation}
\label{eqnn}
\sqrt{n_en_H(r)} = \biggl(\frac{\rho_{g_0}}{1.244m_p}\biggr)[1 +
(r/r_{core})^2]^{-3\beta/2}h^{1/2}.
\end{equation}
Here, $\rho_{g_0}$ is the central gas density, which can be described by
\begin{equation}
\label{eqr0}
\rho_{g_0}=\biggl[\frac{S_0 4\pi(1.244m_p)^2\Gamma(3\beta)(1+z)^4}
{\epsilon_0\times10^{-23}\sqrt{\pi}r_c\Gamma(3\beta-1/2)}\biggr]^{1/2},
\end{equation}
where $\epsilon_0$ is the X-ray volume emissivity coefficient calculated from the most recent
Raymond-Smith plasma code\footnote{\raggedright{Updated 21 Sep., 1993,  available at
ftp://heasarc.gsfc.nasa.gov/software/plasma\_codes/raymond}}.
We combine this with the gas mass density equation (see \cite{car97}1997)
\begin{equation}
\label{eqrg}
\rho_{gas}(r)=1.244m_p\sqrt{n_en_H(r)}h^{1/2}\rm{~gm~cm}^{-3},
\end{equation}
and integrate out to the 2~$\sigma$ cutoff radius to obtain gas masses complementary to the
total X-ray cluster masses.  These are listed as a fraction of total cluster mass in Table
\ref{tablefour.tex}, column (4).  We have overlaid the radial gas mass profiles on Figure
\ref{figmp} (dotted lines). 

For those clusters with suspected cooling flows, we have also estimated the fraction of each
cluster's luminosity attributed to the cooling flow.  We measure the total luminosity inside
the inner cutoff radius described in \S \ref{secanal}, and subtract from that the
luminosity in the same region we calculate from our fitted
$\beta$ model profile. The net luminosity is then expressed as a fraction of the total cluster
luminosity in column (5) of Table \ref{tablefour.tex}.

Under the assumption that rich galaxy clusters have collapsed gravitationally from regions of
the universe with representative ratios of dark to baryonic matter, we can infer that the
baryon fraction in clusters is equal to that in the field.  Neglecting the galaxy mass as
small compared to the X-ray emitting gas mass (see \cite{dav95}), we equate the baryon
fraction with the gas mass fraction. We can then obtain 
$\Omega_0=\Omega_{\rm{b}}/f_{\rm{b}}$ where $\Omega_{\rm{b}}$ is the mass fraction of the
critical density in baryons, and $f_b$ is the measured mass fraction of the universe in
baryons. Values of $\Omega_{\rm{b}}$ can be constrained by Big Bang nucleosynthesis models
(c.f. \cite{whi93}) as well as light element abundance measurements. Using D/H ratios,
\cite{tyt98} finds $\Omega_{\rm{b}}h^2 = 0.0193\pm0.0014$. A recent review by
\cite{oli97} suggests the range $0.005<\Omega_{\rm{b}}h^2<0.015$ with the most
likely value of 0.0065. 

From Table \ref{tablefour.tex}, the average gas mass fraction for our cluster sample is
$0.047\pm0.002h^{-3/2}$, with the uncertainty corresponding to the  weighted average
uncertainty of the gas fraction among the cluster sample. This is in agreement with typical
observed values of ($0.03-0.20h^{-3/2}$, e.g \cite{sar88}; \cite{boh94}; \cite{whi95}).
Combining the gas mass fraction with the \cite{tyt98} range for $\Omega_{\rm{b}}$ yields
\begin{equation}
\label{eqo1}
\Omega_0=0.41\pm0.03 h^{-1/2}.
\end{equation}
For the wider range suggested by Olive, we find
\begin{equation}
\label{eqo2}
\Omega_0=0.14^{+0.18}_{-0.03} h^{-1/2}.
\end{equation}
We emphasize that the baryon fraction is only measured out to a limited radius within the
cluster (the 2 $\sigma$ cutoff), but as can be seen from the mass profiles, the fraction is
increasing with radius.   With accurate data out to greater radii, we would expect to find
somewhat larger baryon fractions, implying that we have actually presented upper limits to
the value of $\Omega_0$. \cite{car96} analyzed the CNOC sample and used cluster M/L ratios to
obtain $\Omega_0=0.19\pm$0.06, in general agreement with this work. The addition of galaxy
masses would increase the cluster baryon fraction slightly, and lower the value of
$\Omega_0$. The primary statistical uncertainty in total mass is the temperature uncertainty,
while for the gas mass it is the uncertainty in $\beta$. It should be noted that to achieve
$\Omega_0=1$ would require a drop in the gas mass relative to the total mass by a factor of
$\sim2-3$, clearly not allowed by the data.

\section{Conclusions \label{secconc}}

We have shown that virial cluster masses are in good agreement with the X-ray derived masses
in this intermediate redshift sample. The comparison shows no systematic bias, with a
weighted average $M_{Dyn}/M_X=1.04\pm0.07$,  and no intrinsic scatter in individual
$M_{Dyn}/M_X$ ratios is seen beyond the
$20-30\%$ observed uncertainties. We do note, however, that our
observed agreement is dependent on our choice of relatively relaxed systems which have
extensive optical and X-ray observations. We have estimated that systematic errors due to
temperature gradients in the ICM, cooling flows, and non-equilibrium states may introduce
overall biases of up to
$+25\%$ and
$-10\%$ on the
$M_{Dyn}/M_X$ ratios. This implies that both detailed dynamical and X-ray mass estimates
reflect the true gravitating mass in clusters.

We find that mass estimates from the literature based on gravitational lensing are
systematically higher than the other two methods, as has been noted by other authors.  
However, our X-ray and dynamical data lack resolution on the smallest scales and cannot
accurately probe the cluster cores where the strongest discrepancy exists between our methods
and strong lensing techniques.  We therefore suggest that systematic errors including mass
clumps along the line of sight, unknown redshifts in the lensed objects, and difficulties in
characterizing the X-ray emission in cluster cores still prevent a  detailed comparison of
techniques in the small radii regime.  The effects of cooling flows and merger events further
complicates these studies (\cite{all97}; \cite{wu98}), although the trend in mass estimates
from these effects is not detected in our sample. We suggest that the smaller but
non-negligible discrepancy seen with weak lensing is most likely due to uncertainty in the
redshift distribution of background galaxies, and inconsistencies in extrapolations along the
mass profiles.

The observed baryonic gas fraction in clusters constrains $\Omega_0$ to be less than
$\sim0.4$, confirming the  severe challenge to closed cosmological models with
$\Lambda=0$ posed by previous efforts.

\acknowledgements 
This research has made use of data obtained through the High Energy
Astrophysics Science Archive Research Center Online Service, provided by the NASA/Goddard
Space Flight Center. Available at http://heasarc.gsfc.nasa.gov/cgi-bin/W3Browse/w3browse.pl.
This work was supported by NASA grant NAG5-3161 and NSF grant AST-9617145. We would like
to thank the anonymous referee for useful suggestions that helped to improve the paper.
ADL wishes to acknowledge Ka Chun Yu for help with figures; and Beth White for her ongoing 
support.

\appendix

\section{Appendix: Notes on Individual Clusters \label{secapp}}

\subsection{Abell 2390}
Abell 2390, at z=0.23, is one the best-studied clusters at higher redshifts. It is extremely
bright in X-rays, providing one of the highest signal to noise ratio surface brightness
profiles in our sample.  There are ROSAT HRI, ROSAT PSPC, and ASCA data available.  There are
multiple arcs and arclets visible that have been used for gravitational lensing studies.  It
has been estimated to have a large ($\sim250h^{-2}$~M$_{\sun}$~yr$^{-1}$) cooling flow (Pierre
\etal 1996; Rizza
\etal 1997).

The HRI surface brightness  profile is relatively smooth - apparently a good candidate for a
simple $\beta$ model fit. The emission is noticeably elliptical.  We see dramatic jumps in
emission near the core in this profile. This is probably the signature of the cooling flow,
and we have only used data outside of 39$h^{-1}$kpc for our fit.

Our X-ray centroid lies at R.A.(1950)$=21^h51^m14\fs1$, decl.(1950)$=17^h27^m33\fs1$, which is
within one arcsecond of the measurements of \cite{pie96} and \cite{all97}.

Our fits to $\beta$ and $r_{core}$ differ with the X-ray analyses of both
\cite{riz97} and \cite{pie96} by up to $\sim$30\% due to details of the analyses, primarily
methods of accounting for the effect of the cooling flow. We note, however, that our mass
estimates are in good agreement with both works.  We see a large discrepancy between our
X-ray analysis and that of \cite{all97}, who find an $r_{core}$ four times smaller than ours. 
Combined with their mean temperature of 14.5 keV, compared with the value of 8.9 keV from
\cite{mus97}, and a large fitted intrinsic $N_{H}$ column, this yields a mass profile
far in excess of our own.  We attribute the difference in
$r_{core}$ partly to our subjective method of accounting for the peaked cooling flow on the
profile but also note that the \cite{all97} $r_{core}$ value is much smaller than values for
typical clusters.  The details of the \cite{all97} spatial analysis are in preparation, and
may shed some light on our differences.

Gravitational lensing estimates for the mass based on weak lensing (\cite{squ96}, with the
mass value calculated by \cite{wu97}) and a strong arc at z=0.913 (discovered by
\cite{pel91} and with the mass calculation performed by Pierre \etal 1996) are higher
than our X-ray mass by a factor of 1.8 and 1.4, at radii of 602 and 92$h^{-1}$ kpc,
respectively. \cite{all97} also report a mass estimate based on the same strong gravitational
arc at a radius of 87$h^{-1}$kpc which is 2.3 times greater than our X-ray derived estimate. 

\subsection{MS~0015.9+1609}

MS~0015.9+1609, (also CL0016+16), is another well-studied, bright cluster with available HRI,
PSPC, ASCA, and lensing data.  At z=0.55, it is the most distant cluster in our sample.  We
see evidence in the surface brightness profile for a cooling flow, and we exclude the inner
31 kpc from the fits.   Our X-ray centroid lies at
R.A.(1950)$=00^h15^m58\fs2$, decl.(1950)$=16^h09^m27\fs1$, which is $\sim 7\arcsec$ south of
the BCG coordinates given by Gioia \& Luppino (1994, G94 hereafter), and $\sim 4\arcsec$ south
of the HRI coordinates given by \cite{neu97}. We note that the HRI image appears to have some
asymmetry near the core, including a knot to the northeast. In addition, this cluster was not
observed in H$\alpha$ by \cite{don92}, making our identification of the peak in the surface brightness profile as
a cooling flow somewhat uncertain.

We note that the independent PSPC fitted values for
$\beta$ and
$r_{core}$ are significantly higher than the HRI fits.  However, upon constraining the PSPC
fit with the HRI fitted value of $r_{core}$, the final fitted $\beta$ value is consistent
with the original HRI $\beta$ value.  The PSPC profile is still very well fit with the final
choice of parameters.

\cite{neu97} have completed a similar X-ray analysis using both HRI and PSPC data, and we
note that our mass value is well within the errors of their result at a radius of 
1.5$h^{-1}$  Mpc. Our gas fraction, converted to their cosmology is
0.26$\pm$0.03$h^{-3/2}_{50}$, in agreement with their quoted range of
0.14-0.32$h^{-3/2}_{50}$.

\cite{sma95} analyzed MS~0015.9+1609 using weak lensing, and arrived at a total mass 1.5
times higher than our X-ray mass, at a radius of 334$h^{-1}$ kpc. Their more recent analysis
(\cite{sma97}) obtains a mass 1.1 times higher than ours at a radius of 226$h^{-1}$ kpc.

\subsection{MS~0302.7+1658}

The relatively low counts in the HRI exposure of this cluster at z=0.43 yield a peaked but
not uniform surface brightness profile, from which we exclude the inner 27$h^{-1}$ kpc,
assuming a cooling flow. Our X-ray centroid lies at
R.A.(1950)$=03^h02^m43\fs3$, decl.(1950)$=16^h58^m29\fs4$, which is within $\sim 2\arcsec$ of
the BCG coordinates given by \cite{gio94}.

This cluster has not been observed by ASCA, and we have used the empirical Bgc-$T_x$ relation
(\cite{yee99}) to estimate the temperature.

\subsection{MS~0440.5+0204}

This cluster lies at z=0.20, and has similarly low HRI counts as MS~0302.7+1658, thus making 
the estimation of a cooling flow radius somewhat difficult. However the combination of the
surface brightness profile and the detection of extended H${\alpha}$ emission (\cite{don92})
lead us to exclude the inner 21 kpc.   Our X-ray centroid lies at
R.A.(1950)$=04^h40^m33\fs8$, decl.(1950)$=02^h04^m44\fs7$, which is within $\sim 1\arcsec$ of
the BCG coordinates given by \cite{gio94}. We note that the fitted gas mass fraction for this
cluster is the lowest in the sample, achieving a value of 0.026$h^{-3/2}$ at 168$h^{-1}$kpc.

Recently, \cite{gio98} analyzed this cluster, and obtained a velocity dispersion of 872 km
s$^{-1}$, 1.4 times higher than that of \cite{car96}. Upon comparison, we find
several common galaxies with identical redshifts, and we can find no a priori reason to
discount either value for the velocity dispersion. Thus we have expanded the uncertainty on
our velocity dispersion to encompass their value.

\cite{gio98} also performed an analysis of the available ASCA and ROSAT data similar to our
own, obtaining significantly smaller values of $r_{core}$ and
$\beta$, as well as total mass (when we compare at a similar radius).  We attribute this to
our exclusion of X-ray emission in the core from the fit to the surface brightness profile.

The \cite{wu97} review contains a mass estimate within  46$h^{-1}$ kpc based on a strong arc
at z=0.53 discovered by \cite{lup93}.  Their value is 2.5 times our X-ray derived mass at the
same radius, in the same cosmology.
\cite{gio98} also placed limits from strong lensing on the cluster mass within 52$h^{-1}$ kpc
which were $\sim1.9$ times higher than our X-ray estimate.

\subsection{MS~0451.5+0250}

MS~0451.5+0250 (also Abell 520) at z=0.20 exhibits extremely disturbed, clumpy morphology in
a relatively short ROSAT HRI exposure, resulting in a surface brightness profile that is not
well fit by a $\beta$ model.  We have therefore chosen to use the PSPC data for our fits,
but we constrain $r_{core}$ to be $100\pm50$ kpc in order to avoid overestimating this
parameter due to the large PSF of the instrument. It is also noted in
\cite{car96} that it contains two dominant galaxies, and two peaks in its galaxy velocity
distribution.  We suggest this as a strong candidate for a core merger. We also identify it
as a non-cooling flow cluster based on the lack of a bright peak in the X-ray surface
brightness profile, and a non-detection of H$\alpha$ by \cite{don92}. Our X-ray centroid (from
the PSPC data) lies at R.A.(1950)$=04^h51^m32\fs4$, decl.(1950)$=02^h50^m18\fs2$, which is
within
$\sim 30\arcsec$ of the optical center given by \cite{gio94}. They note, ``There is not a well
defined optical center in this cluster but several galaxy condensations are visible.''

Interestingly, the PSPC exposure yields a reasonably smooth surface brightness
profile, which is well fit by a $\beta$ model.  We anticipate that future imaging of this
cluster with higher resolution instrumentation (e.g. AXAF, XMT) may reveal new insight into
merging cluster structure.

\subsection{MS~0451.5-0305}

The analysis of the PSPC image of this  z=0.54 cluster by \cite{don96} revealed one of the
first extremely hot, high redshift clusters ever found.  There is no evidence of a cooling
flow present. Our X-ray centroid lies at
R.A.(1950)$=04^h51^m41\fs3$, decl.(1950)$=-03^h05^m54\fs8$, which is within
$\sim 15\arcsec$ of the optical center given by \cite{gio94}. We attribute this relatively
large offset to the non-uniform core of the X-ray image, containing multiple peaks, and
exhibiting elongation both N-S and E-W.

The fitted parameter values are markedly different from HRI and PSPC data sets (see Table
\ref{tabletwo.tex}).  The combination of the two results in an intermediate value of $\beta$
which fits both the HRI and the PSPC profiles extremely well.  We note that the use of the
final parameters from this work results in a total mass at a radius of 1$h^{-1}$Mpc which is
30\% lower than that derived from PSPC data alone by \cite{don96}, though within the somewhat
large errors from that work. We have assumed the temperature calculated by Donahue for our
analysis.

Strong arc lensing mass estimates from \cite{lup95} and \cite{wu97} (based on an arc from the
\cite{lef94} survey) find values in excess of our X-ray masses by factors of 2.3 and 3.5, at
radii of 132 and  106$h^{-1}$ kpc, respectively.  Redshifts of the lensed arcs are not known.

\subsection{MS~0839.8+2938}

For the otherwise average cluster MS~0839.8+2938, at z=0.19,  we estimated a cooling flow
cutoff radius of 25 kpc.  Our cooling flow identification is supported by the detection of
extended H$\alpha$ emission by \cite{don92}. Our X-ray centroid lies at
R.A.(1950)$=08^h39^m53\fs2$, decl.(1950)$=29^h38^m13\fs7$, which is within
$\sim 5\arcsec$ of the optical center given by \cite{gio94}.

\subsection {MS~0906.5+1110}

This cluster at z=0.17 exhibits a bimodal galaxy velocity distribution, suggesting a recent
merger.  The X-ray emission exhibits a significant secondary peak near the center.  The
surface brightness profile is irregular within the core radius, but has a very smooth tail.
We also identify it
as a non-cooling flow cluster based on the non-detection of H$\alpha$ by \cite{don92}. A
detailed substructure analysis may reveal a correlation between the X-ray morphology and
galaxy colors and/or velocities.  Our X-ray centroid lies at R.A.(1950)$=09^h06^m29\fs8$,
decl.(1950)$=11^h10^m43\fs2$, which is within
$\sim 5\arcsec$ of the optical center given by \cite{gio94}.

This cluster has not been observed by ASCA, and we have used the L$_x$--T$_x$ relation from
\cite{mus97} to obtain a temperature. Because of the spuriously high velocity dispersion found
for the galaxies (\cite{car96}), we have omitted this cluster from our comparison of
dynamical and X-ray mass estimates.

\subsection{MS~1006.0+1202}

The long pointed exposure of this relatively faint cluster at z=0.26 reveals a fair amount of
substructure and asymmetry in the HRI image.  The resulting surface brightness profile is
flattened, yielding the lowest fitted value of $\beta$ in the sample.  There is possible
evidence for a very small cooling flow, though it is not definite; there is a weak detection
of extended H$\alpha$ from \cite{don92}. We exclude only the inner 4$h^{-1}$ kpc from the
fit.  This would be another interesting candidate for further study.  Our X-ray centroid lies
at R.A.(1950)$=10^h06^m07\fs4$, decl.(1950)$=12^h02^m23\fs8$, which is within
$\sim 5\arcsec$ of the optical center given by \cite{gio94}.

This cluster has not been observed by ASCA, and we have used the Bgc-$T_x$ relation to
estimate a temperature.

Three strong arc mass estimates by \cite{wu97} (based on the \cite{lef94} survey)
have been compared with this work, yielding ratios of 3.5, 3.8 and 4.6, at radii of  42, 74
and 147$h^{-1}$ kpc,  respectively.  The arc redshifts are not known.  The extremely high
discrepancy here reinforces our opinion that comparison between strong arcs and X-ray mass
estimates are not always applicable. Because of the small radius involved, the lensing method
is particularly sensitive to the mass distribution model assumed, and can apparently be
grossly  different from the X-ray derived mass profile.  The X-ray mass, in turn, must be
extrapolated deep into the core, and cannot account for irregular mass distributions at very
small scales which have little power in the overall profile.

\subsection{MS~1008.1-1224}

An extremely regular cluster in most respects, MS~1008.1-1224, at z=0.31 has a
characteristically smooth surface brightness profile, no evidence of cooling structure, and
mild asymmetry in the HRI image, consisting of faint extended emission to the North. We also
identify this cluster as a non-cooling flow based on the non-detection of H$\alpha$ by
\cite{don92}.Our X-ray centroid lies at R.A.(1950)$=10^h08^m05\fs4$,
decl.(1950)$=-12^h24^m44\fs6$, which is
$\sim 15\arcsec$ north of the optical center given by \cite{gio94}, who note ``The cluster
has a circular distribution of galaxies surrounding a NS-elongated core. There is a secondary
clump of galaxies to the N.'' We expect that the northern X-ray emission accounts for the
difference in centroids.

The strong arc mass estimate of \cite{wu97} (based on a \cite{lef94} arc) is greater than the
X-ray mass by a factor of 3.7 at a radius of  138$h^{-1}$kpc. The redshift of the arc is not
known.

\subsection{MS~1224.7+2007}

The HRI image of MS~1224.7+2007, at z=0.33 exhibits a strong, radially symmetric core with
extremely faint and ragged edges.  The surface brightness profile is enigmatic.  After
extensive investigation, an inner cooling flow cutoff radius of 34 kpc was chosen.  The
extended H${\alpha}$ emission detected by \cite{don92} from this cluster is consistent with a
fairly large cooling flow. Our X-ray centroid lies at R.A.(1950)$=12^h24^m42\fs6$,
decl.(1950)$=20^h07^m31\fs4$, which is within
$\sim 2\arcsec$ of the optical center given by \cite{gio94}.

Fahlman \etal (1994) have performed a weak lensing analysis of MS~1224.7+2007, and their mass
estimate is 1.7 times greater than ours at a radius of 512$h^{-1}$kpc. 

\subsection{MS~1358.4+6245}

MS~1358.4+6245, at z=0.33 exhibits an extremely regular surface brightness profile from its
PSPC image, and a fairly disturbed profile from HRI data.  The galaxy velocity distribution
observed by C96 indicates that there is a small common velocity group south of the optical
cluster center. Our X-ray centroid lies at
R.A.(1950)$=13^h58^m20\fs6$, decl.(1950)$=62^h45^m33\fs3$, which is within
$\sim 5\arcsec$ of the optical center given by \cite{gio94}. To account for a cooling flow
apparent in both the HRI image and the H$\alpha$ data from \cite{don92}, an inner cutoff
radius of 36 kpc was adopted.

This cluster also contains a strongly lensed arc with the highest measured redshift known,
z=4.92 (\cite{fra97}). \cite{all97} have performed a strong lensing analysis of
MS~1358.4+6245, obtaining a mass 1.7 times greater than ours at a radius of 65$h^{-1}$kpc.
They also perform an X-ray deprojection analysis, obtaining a mass 3.8 times higher than ours
at 65$h^{-1}$kpc.   The mean temperature in our analyses differs by only $\sim$15\%. 
However, the Allen fitted value for
$r_{core}$ is 5.2 times smaller than our fitted value.  The HRI surface brightness profile of
MS~1358.4+6245 indicates a very peaked core, with significant irregularity directly around
it, followed by a reasonably smooth tail.  Because the strong lensing probes only the very
core of the cluster, we may be seeing very pronounced effects due to subclumping, which our
cooling flow cutoff method does not account for.  We emphasize, however, that the well
measured dynamical mass is in excellent agreement with our X-ray derived mass at a radius of
240$h^{-1}$kpc.  In this case, our X-ray analysis may only provide strong constraints on the
mass distribution at larger radii.

\subsection{MS~1455.0+2232}

MS~1455.0+2232 at z=0.26, displays the highest signal to noise ratio HRI surface brightness
profile in this sample after Abell 2390.  It is highly radially symmetric, and strongly
peaked.   This cluster has the highest H$\alpha$ luminosity and the largest H$\alpha$  extent
among the \cite{don92} sample. Allen \etal (1996) estimate it to have one of the largest
cooling flows known,
$\sim$1500 M$_{\sun}$yr$^{-1}$.  Our X-ray centroid lies at R.A.(1950)$=14^h55^m00\fs4$,
decl.(1950)$=22^h32^m36\fs0$, which is within
$\sim 5\arcsec$ of the optical center given by \cite{gio94}.  The HRI surface brightness
profile is extremely well constrained by the slope of the profile, and the values of the
fitted parameters are insensitive to choice of inner cutoff radius out to a value of 103 kpc.
Therefore, based on the surface brightness profile, we have chosen a cutoff of only 20
kpc,  which provides a good statistical fit to the HRI data.

Our mass estimate is in close agreement with the X-ray derived mass estimate of Allen \etal
(1996), differing by only 5\%.  The gravitational lensing mass estimates, based on a
strong arc of unknown redshift at 52$h^{-1}$kpc (\cite{wu97}, from an arc in the \cite{lef94}
survey) and weak  lensing observed by \cite{sma95} at  237$h^{-1}$kpc, are larger by factors
of 1.8 and 1.3, respectively.

\subsection{MS~1512.4+3647}

The HRI image of MS~1512.4+3647, at z=0.37 is somewhat elliptical, uniform, and mildly
peaked.  There is evidence for a small cooling flow, and the parameter fits are only mildly
sensitive to inner cutoff radius. There is a weak detection of extended H$\alpha$ emission
by \cite{don92}. We have excluded the inner 11$h^{-1}$ kpc from the fits.  Our X-ray centroid
lies at R.A.(1950)$=15^h12^m26\fs0$, decl.(1950)$=36^h47^m24\fs7$, which is within
$\sim 5\arcsec$ of the optical center given by \cite{gio94}.

\cite{ham97} perform a similar X-ray analysis, yielding a total mass within 315$h^{-1}$kpc
which is 3\% less than our own.  We note that their gas mass fraction of 20\% is a factor of
$\sim$3 higher than our estimate. The difference between our analyses may hinge on the
neutral hydrogen absorption of the cluster, which we assume to be the  value from
\cite{dicwww}, and which they  leave as a free parameter in their fit.

\cite{ham97} also make use of a PSPC image to supplement their analysis.  We have chosen not
to include these data, due to the  brevity of the PSPC exposure, and the fact that the cluster
is not centered in the field of view, making accurate flat fielding even more difficult. Note
that the inclusion of PSPC data does not significantly change their parameter fits, yielding
a total mass in close agreement with our result.

\clearpage

\begin{figure}
\caption{HRI Images of the CNOC clusters with X-ray contours overlaid. Images have been
blocked by a factor of 8, and smoothed with a 1 pixel FWHM gaussian.
The dashed circles represent the 2~$\sigma$ radius for each cluster,
defined in \S \ref{secdis}. These circles are omitted for MS~1231.3+1542 and MS~1621.5+2640
 as the data were insufficient for analysis of the surface brightness profiles.
For Abell~2390, the circle extends beyond the boundary of the figure. Contours are in units of
counts~s$^{-1}$~arcmin$^{-2}$. For Abell~2390 the contours start at 0.05,
and increase in linear steps of 0.05. MS~0015.9+1609: lowest contour at
0.025 with steps of 0.01. MS~0302.7+1658: lowest contour at 0.024 with steps
of 0.003 MS~0440.5+0204: lowest contour at 0.03 with steps of 0.015.
MS~0451.5+0250: lowest contour at 0.04 with steps of 0.01. MS~0451.6-0305:
lowest contour at 0.03 with steps of 0.008. MS~0839.8+2938: lowest contour
at 0.04 with steps of 0.03. MS~0906.5+1110: lowest contour at 0.035 with
steps of 0.01. MS~1006.0+1202: lowest contour at 0.035 with steps of 0.005.
MS~1008.1-1224: lowest contour at 0.03 with steps of 0.008.
MS~1224.7+2007: lowest contour at 0.035 with steps of 0.01.
MS~1231.3+1542: lowest contour at 0.033 with steps of 0.003.
MS~1358.4+6245: lowest contour at 0.04 with steps of 0.02.
MS~1455.0+2232: lowest contour at 0.05 with steps of 0.1.
MS~1512.4+3647: lowest contour at 0.03 with steps of 0.02
MS~1621.5+2640: lowest contour at 0.027 with steps of 0.003.
\label{figpp}}
\end{figure}
\begin{figure}
\caption{X-ray surface brightness profiles with
$\beta$-model fits overlaid.  The solid curve corresponds to the fit when all data points are
included.  The dotted curve (where applicable) is the fit when inner data points have been
excluded from the fit based on our subjective criterion for a cooling flow (see section
\ref{secredux}).   The solid vertical line indicates the radius at which the fitted profile
drops to 2 $\sigma$ above the measured background noise. The dash-dotted vertical line
indicates (where applicable) the inner cutoff radius for the suspected cooling flow.
Horizontal lines correspond to the measured mean background value.
\label{figsb}}
\end{figure}
\begin{figure}
\caption{H$\alpha$/X-ray Luminosity Ratio vs. Cooling Flow Cutoff radius. The y-axis is the
ratio of H$\alpha$ luminosity (from  Donahue \etal 1992) to X-ray Luminosity in the {\it
Einstein}  IPC 0.3--3.5 keV band.  The x-axis indicates the radius, in kpc,  of the
subjectively defined inner cooling flow cutoff.
\label{figha}}
\end{figure}
\begin{figure}
\caption{Radial mass profiles for the cluster sample. The solid line
represents the total mass distribution derived from the X-ray data,  the dashed line the mass
distribution derived from dynamical data, and the dotted line represents the X-ray derived
gas mass distribution. Vertical lines have the same meaning as in Figure \ref{figsb}.
\label{figmp}}
\end{figure}
\begin{figure}
\caption{$M_{Dyn}/M_X$ ratio vs. X-ray derived mass within the HRI 2 $\sigma$ radius.  Errors
on each ratio include the 1~$\sigma$ errors from the $\beta$ model fit and the X-ray
temperatures (see Table \ref{tabletwo.tex}) as well as the uncertainties from the dynamical
mass estimates.
\label{figdmm}}
\end{figure}
\begin{figure}
\caption{$M_{Dyn}/M_X$ ratio vs. the HRI 2 $\sigma$ radius at which $M_X$ is measured for
each cluster. Errors are the same as in figure
\ref{figdmm}.
\label{figdmr}}
\end{figure}
\begin{figure}
\caption{Gravitational lens mass/X-ray derived mass ratio vs. radius. The y-axis is the ratio
of the published gravitational lensing mass estimates to our X-ray derived mass at the same
radius, and in the same cosmology. The x-axis is the radius of the lensing mass estimate in
our cosmology  ($h=100$, $q_0=0.1$).  Note that some clusters have gravitational lens mass
estimates at multiple radii and hence are represented by more than one data point.  Error
bars on the mass ratios include both the lensing mass error and  the X-ray derived mass
error, and are only plotted where a gravitational lens  mass error was available.  For
reference, error bars for the X-ray masses  are indicated in Figure 
\ref{figdmm}.    Filled triangles represent strong lensing observations where the redshift of
the lensed arc was known, unfilled triangles represent strong lensing of  an arc of unknown
redshift, and filled squares represent weak lensing  analyses. To the right of each data
point is the cluster name. To the  left of each data point is the reference for the lensing
data:  (1) Wu \& Fang 1997; (2) Pierre \etal 1996; (3) Smail \etal 1995;  (4) Smail \etal
1997; (5) Luppino \& Gioia 1995; (6) Fahlman \etal 1994; (7) A98; (8) Gioia \etal 1998.
\label{figgmr}}
\end{figure}
\voffset=-1.0in
\begin{deluxetable}{lrrcc}
\tablecaption{Cluster Sample \label{tableone.tex}}
\tablewidth{0pt}
\tablehead{
\colhead{Cluster} & \colhead{$z$} & \colhead{$L_x$\tablenotemark{a}} &
\colhead{HRI Exposure\tablenotemark{b}} & \colhead{PSPC Exposure\tablenotemark{b}} }
\startdata
Abell 2390 & 0.2279 & 5.5 & 32904 & 9287 \nl
MS 0015.9+1609 & 0.5466 & 14.6 & 73333 & 41405 \nl
MS 0302.7+1658 & 0.4246 & 5.0 & 32706 & \nodata \nl
MS 0440.5+0204 & 0.1965 & 4.0 & 26333 & \nodata \nl
MS 0451.5+0250 & 0.2010 & 7.0 & 12084 & 4743  \nl
MS 0451.6-0305 & 0.5392 & 20.0 & 45371 & 15296  \nl
MS 0839.8+2938 & 0.1928 & 5.4 & 18797 & \nodata \nl
MS 0906.5+1110 & 0.1709 & 5.8 & 78218 & \nodata \nl
MS 1006.0+1202 & 0.2605 & 4.8 & 79903 & \nodata \nl
MS 1008.1-1224 & 0.3062 & 4.5 & 66780 & \nodata \nl
MS 1224.7+2007 & 0.3255 & 4.6 & 52256 & \nodata \nl
MS 1231.3+1542 & 0.2350 & 2.9 & 62967 & 1246 \nl
MS 1358.4+6245 & 0.3290 & 10.6 & 28858 & 21709 \nl
MS 1455.0+2232 & 0.2570 & 16.0 & 14703 & \nodata \nl
MS 1512.4+3647 & 0.3726 & 4.8 & 34035 & 5230 \nl
MS 1621.5+2640 & 0.4274	& 4.5 & 43256 & \nodata \nl
\enddata
\tablenotetext{a}{$10^{44}$~erg~s$^{-1}$ in the {\it Einstein} IPC band ($0.3-3.5$~keV).} 
\tablenotetext{b}{Corrected exposure time in seconds (see text for corrections applied).}
\end{deluxetable}
\small{
\hoffset=-0.5in
\begin{deluxetable}{lrcrcrccc}
\tablecaption{Fitted $\beta$ Model Parameters \label{tabletwo.tex}}
\tablewidth{0pt}
\tablehead{
\colhead{Cluster} & \colhead{HRI $\beta$\tablenotemark{a}} & \colhead{HRI
$r_{core}$\tablenotemark{a}} & \colhead{PSPC $\beta$\tablenotemark{b}} & \colhead{PSPC
$r_{core}$\tablenotemark{b}} & \colhead{Final $\beta$\tablenotemark{c}} & \colhead{Final
$r_{core}$\tablenotemark{c}} & \colhead{Temperature\tablenotemark{d}} &
\colhead{Ref.\tablenotemark{e}} }
\startdata
Abell 2390 & 0.61$\pm$0.02 & 100$\pm$7 & 0.69$\pm$0.03 & 138$\pm$13 & 0.62$^{+0.01}_{-0.01}$ &
103$^{+4}_{-10}$ & $8.9^{+0.5}_{-0.4}$ & 1 \nl 

MS 0015.9+1609 & 0.87$\pm$0.08 & 214$\pm$24 & 0.90$\pm$0.04 & 255$\pm$16 & 0.81$^{+0.02}_{-0.08}$ &
196$^{+5}_{-23}$ & $8.0^{+0.5}_{-0.5}$ & 1 \nl

MS 0302.7+1658 & 0.68$\pm$0.15 & 35$\pm$19 & \nodata & \nodata & 0.68$\pm$0.15 & 35$\pm$19 &
$3.8^{+1.6}_{-1.6}$ & 5 \nl

MS 0440.5+0204 & 0.68$\pm$0.10 & 54$\pm$16 &\nodata & \nodata & 0.68$^{+0.10}_{-0.10}$ &
54$^{+16}_{-16}$ &
$5.3^{+0.6}_{-0.4}$ & 1 \nl

MS 0451.5+0250 & 0.80$\pm$11.1 & 235$\pm$926 & 0.49$^{+0.08}_{-0.08}$ & 100$^{+50}_{-50}$ &
0.49$^{+0.08}_{-0.08}$ & 100$^{+50}_{-50}$ & $8.6^{+0.5}_{-0.5}$ & 1 \nl

MS 0451.6-0305 & 0.80$\pm$0.07 & 165$\pm$18 & 0.92$\pm$0.11 & 189$\pm$30 & 0.84$^{+0.06}_{-0.06}$ &
175$^{+14}_{-15}$ & $10.4^{+0.8}_{-0.7}$ & 2 \nl

MS 0839.8+2938 & 0.81$\pm$0.16 & 69$\pm$20 &\nodata & \nodata & 0.81$\pm$0.16 & 69$\pm$20 &
$4.2^{+0.2}_{-0.2}$ & 1 \nl

MS 0906.5+1110 & 0.93$\pm$0.07 & 147$\pm$11 & \nodata & \nodata & 0.93$\pm$0.07 & 147$\pm$11 &
$8.0^{+2.0}_{-1.0}$ & 3  \nl

MS 1006.0+1202 & 0.67$\pm$0.05 & 165$\pm$18 & \nodata & \nodata & 0.67$\pm$0.05 & 165$\pm$18 &
$7.2^{+1.6}_{-1.6}$ & 5  \nl

MS 1008.1-1224 & 0.68$\pm$0.05 & 124$\pm$14 & \nodata & \nodata & 0.68$\pm$0.05 & 124$\pm$14 &
$7.3^{+1.2}_{-0.8}$ & 1 \nl

MS 1224.7+2007 & 0.58$\pm$0.10 & 63$\pm$29 & \nodata & \nodata & 0.58$\pm$0.10 & 63$\pm$29 &
$4.3^{+0.7}_{-0.6}$ & 4 \nl

MS 1358.4+6245 & 0.84$\pm$0.15 & 137$\pm$31 & 0.75$\pm$0.03 & 107$\pm$8 & 0.83$^{+0.11}_{-0.10}$ &
133$^{+21}_{-20}$ & $6.5^{+0.3}_{-0.3}$ & 1 \nl

MS 1455.0+2232 & 0.66$\pm0.03$ & 40$\pm$4 & \nodata & \nodata & 0.66$\pm0.03$ & 40$\pm$4 &
$5.5^{+0.2}_{-0.1}$ & 1 \nl

MS 1512.4+3647 & 0.55$\pm$0.04 & 29$\pm$7 & \nodata & \nodata & 0.55$\pm$0.04 & 29$\pm$7 &
$3.6^{+0.7}_{-0.4}$ & 1 \nl

\tableline
Average Values & 0.72$^{+0.11}_{-0.08}$ & 103$\pm$57 & 0.75$\pm$0.16 & 158$\pm$58 &
0.72$\pm$0.08 & 103$\pm$58 & 6.5$\pm$2.1 \nl
\enddata
\tablenotetext{a}{Independent HRI data, $r_{core}$ in $h^{-1}$~kpc} 
\tablenotetext{b}{Independent PSPC data, $r_{core}$ in $h^{-1}$~kpc}
\tablenotetext{c}{Where PSPC data are available, these parameters are fit with combined data, using
the HRI $r_{core}$ value to retrieve a fitted
$\beta$ value from the PSPC profile.  For MS 0451.5+0250 the
independent PSPC data is used with a fixed core radius of 100$\pm$50~kpc  - see Appendix.}
\tablenotetext{d}{Mean ASCA cluster Temperature in keV, except as noted.}
\tablenotetext{e}{Source for Temperature value.}
\tablecomments{Uncertainties in all columns are 1~$\sigma$.  MS 0451.5+0250 has been excluded
from the average calculation of HRI parameters.}
\tablerefs{(1) Mushotzky \& Scharf 1997; (2) Donahue 1996; (3) L$_x$--T$_x$ relation,
e.g. Mushotzky \& Scharf 1997; (4) Henry 1997, (5) Bgc-T$_x$ relation, Yee \etal 1999}
\end{deluxetable}
\normalsize{
\hoffset=-0.25in
\begin{deluxetable}{lcccccc}
\tablecaption{Cluster Masses \label{tablethree.tex}}
\tablewidth{0pt}
\tablehead{
\colhead{Cluster\tablenotemark{a}} & \colhead{Inner Cutoff} &\colhead{HRI 2 $\sigma$} &
\colhead{X-ray Mass} &
\colhead{Dynamical/X-ray} & \colhead{PSPC 2 $\sigma$} & \colhead{X-ray Mass} \nl
& \colhead{Radius} & \colhead{Radius} & \colhead{$10^{14}h^{-1}M_{\sun}$} & \colhead{Mass
Ratio} & \colhead{Radius} & \colhead{$10^{14}h^{-1}M_{\sun}$} \nl 
& \colhead{$h^{-1}$ kpc} & \colhead{$h^{-1}$ kpc} & \colhead{(HRI 2 $\sigma$)} &
\colhead{(HRI 2 $\sigma$)} & 
\colhead{$h^{-1}$ kpc} & \colhead{(PSPC 2 $\sigma$)}
}
\startdata
Abell 2390 & 39 & 526 & 3.17$^{+0.18}_{-0.15}$ & 1.09$^{+0.13}_{-0.13}$ & 1283 &
8.40$\pm$0.49  \nl

MS 0015.9+1609* & 31 & 482 & 3.02$^{+0.20}_{-0.36}$ & 1.30$^{+0.26}_{-0.29}$ &
1317 & 8.67$\pm$0.60 \nl

MS 0302.7+1658 & 27 & 141 & 0.39$^{+0.19}_{-0.19}$ & 0.69$^{+0.40}_{-0.40}$ & \nodata &
\nodata \nl

MS 0440.5+0204 & 21 & 168 & 0.62$^{+0.12}_{-0.10}$ & 0.47$^{+0.29}_{-0.12}$ & \nodata &
\nodata \nl

MS 0451.5+0250* & \nodata & 400 & 1.81$^{+0.32}_{-0.31}$ &
1.23$^{+0.29}_{-0.29}$ & 1170 & 7.50$\pm$0.70 \nl

MS 0451.6-0305* & \nodata & 402 & 3.34$^{+0.35}_{-0.31}$ &
1.22$^{+0.24}_{-0.23}$ & 761 & 6.62$\pm$0.58  \nl

MS 0839.8+2938 & 25 & 170 & 0.56$^{+0.11}_{-0.11}$ & 0.68$^{+0.26}_{-0.26}$ & \nodata &
\nodata  \nl

MS 0906.5+1110* & \nodata & 300 & 2.04$^{+0.53}_{-0.30}$ & \nodata & \nodata &
\nodata \nl

MS 1006.0+1202* & 4 & 459 & 2.23$^{+0.55}_{-0.55}$ & 0.98$^{+0.30}_{-0.30}$ &
\nodata &
\nodata \nl

MS 1008.1-1224* & \nodata & 361 & 1.82$^{+0.34}_{-0.23}$ &
1.13$^{+0.28}_{-0.23}$ & \nodata &
\nodata
\nl

MS 1224.7+2007 & 30 & 199 & 0.51$^{+0.12}_{-0.12}$ & 1.20$^{+0.51}_{-0.51}$ & \nodata &
\nodata  \nl

MS 1358.4+6245 & 36 & 286 & 1.46$^{+0.21}_{-0.19}$ & 0.93$^{+0.18}_{-0.16}$ & 764 &
4.09$\pm$0.38  \nl

MS 1455.0+2232 & 20 & 260 & 1.03$^{+0.06}_{-0.05}$ & 1.34$^{+0.26}_{-0.26}$ & \nodata &
\nodata  \nl

MS 1512.4+3647 & 11 & 237 & 0.52$^{+0.11}_{-0.07}$ & 1.17$^{+0.59}_{-0.55}$ & \nodata &
\nodata  \nl

\tableline
Weighted Average & & \nodata & \nodata  & 1.04$\pm$0.07 & \nodata &  \nodata  \nl
\enddata
\tablenotetext{a}{Clusters exhibiting obvious substructure or asymmetry in
their HRI images are denoted with an asterisk.}
\tablecomments{Uncertainties for all masses are 1 $\sigma$.   Uncertainty in the average
value is the uncertainty in the mean. MS 0906.5+1110 has been excluded from the
dynamical/X-ray mass ratio average because the optical data does not constrain
the virial mass.}
\end{deluxetable}
\hoffset=-0.35in
\begin{deluxetable}{lcccc}
\tablecaption{Flux, Luminosity, Gas Fractions, Cooling Flow Luminosity Fractions
\label{tablefour.tex}}
\tablewidth{0pt}
\tablehead{
\colhead{Cluster} & \colhead{HRI $f_x$ $\times$10$^{-12}$} & \colhead{HRI
$L_x \times$10$^{44}$} &
\colhead{Gas Mass Fraction} & \colhead{Cooling Flow}\nl 
& \colhead{erg s$^{-1}$ cm$^{-2}$} & \colhead{erg s$^{-1}$} &
\colhead{HRI 2-$\sigma$ Radius} & \colhead{Luminosity Fraction\tablenotemark{c}} \nl 
& \colhead{$0.1-2.4$~keV\tablenotemark{a}} & \colhead{$0.1-2.4$~keV\tablenotemark{b}} &
\colhead{$h^{-3/2}$} } 
\startdata
Abell 2390 & 13.72 & 11.87 & 0.089$\pm$0.008 & 0.075$\pm$0.003 \nl 
MS 0015.9+1609 & 2.72 & 14.00 & 0.092$\pm$0.010 & 0.018$\pm$0.002 \nl 
MS 0302.7+1658 & 0.39 & 1.63 & 0.039$\pm$0.088 & 0.000$\pm$0.000 \nl 
MS 0440.5+0204 & 1.61 & 1.17 & 0.026$\pm$0.024 & 0.088$\pm$0.011 \nl 
MS 0451.5+0250 & 8.77 & 6.05 & 0.074$\pm$0.012 & \nodata \nl 
MS 0451.6-0305 & 2.39 & 12.01 & 0.067$\pm$0.008 & \nodata \nl 
MS 0839.8+2938 & 2.81 & 17.04 & 0.033$\pm$0.020 & 0.141$\pm$0.013 \nl 
MS 0906.5+1110 & 6.76 & 3.09 & 0.028$\pm$0.003 & \nodata \nl 
MS 1006.0+1202 & 6.00 & 6.35 & 0.064$\pm$0.009 & 0.001$\pm$0.001 \nl 
MS 1008.1-1224 & 3.22 & 5.37 & 0.054$\pm$0.009 & \nodata \nl 
MS 1224.7+2007 & 9.19 & 1.58 & 0.037$\pm$0.245 & 0.187$\pm$0.023 \nl 
MS 1358.4+6245 & 2.42 & 3.86 & 0.047$\pm$0.010 & 0.138$\pm$0.012 \nl 
MS 1455.0+2232 & 6.70 & 6.91 & 0.076$\pm$0.016 & 0.054$\pm$0.004 \nl 
MS 1512.4+3647 & 1.15 & 2.47 & 0.069$\pm$0.032 & 0.019$\pm$0.004 \nl 
\tableline
Average Values & \nodata & \nodata & 0.047$\pm$0.002 & \nodata \nl 
\enddata
\tablenotetext{a}{The flux calculated includes absorption by
Galactic hydrogen. $N_H$ values are taken from \cite{dicwww}.}
\tablenotetext{b}{The luminosity is calculated in the rest frame of the cluster.}
\tablenotetext{c}{The luminosity fraction is calculated at the Inner Cutoff Radius,
listed in  Table \ref{tablethree.tex}.}
\tablecomments{The gas mass fraction average is a weighted average.  The error on the gas
mass fraction average value is the rms standard deviation.}  
\end{deluxetable}
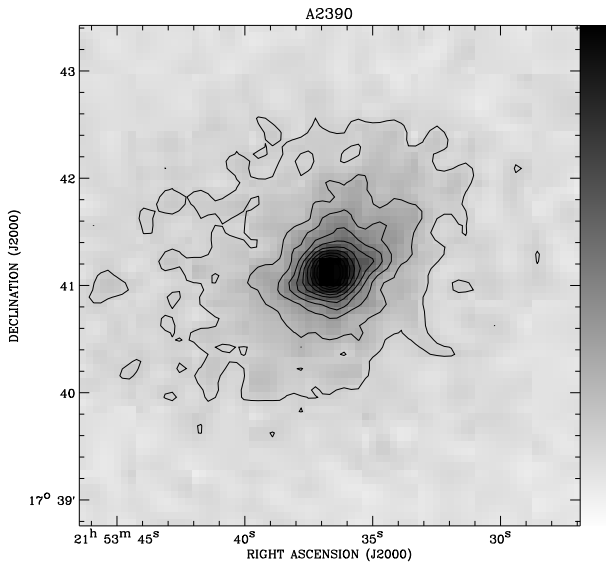
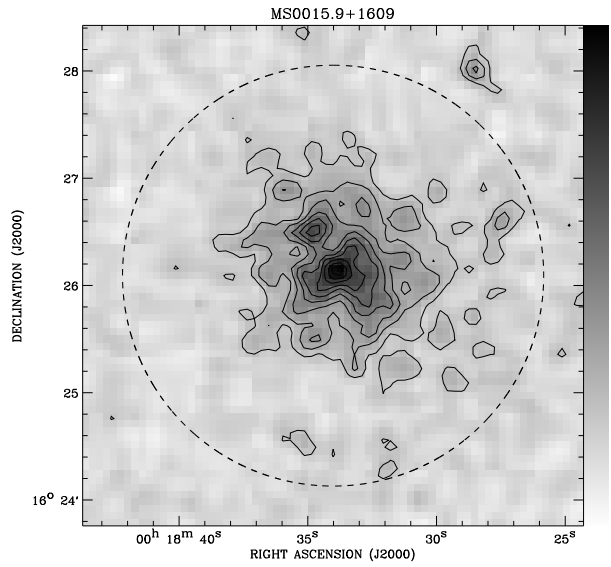
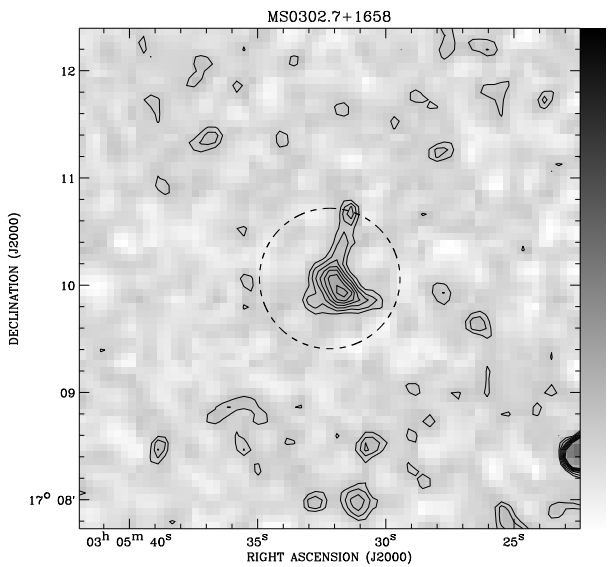
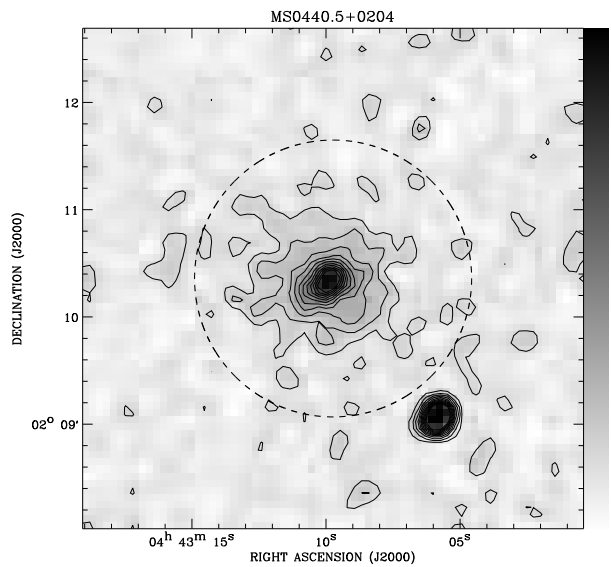
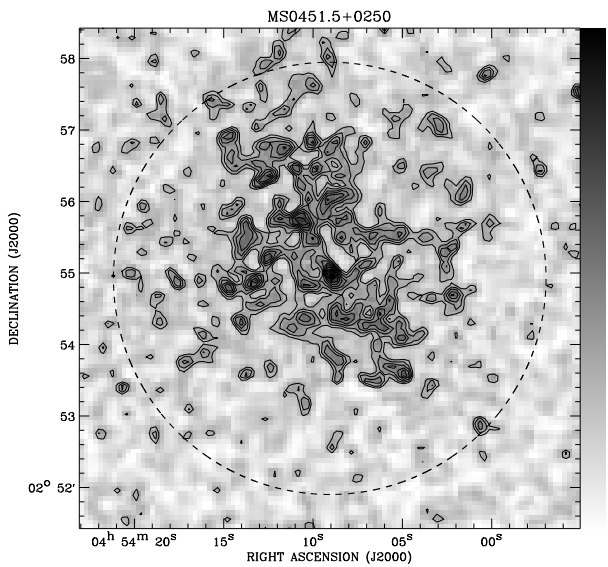
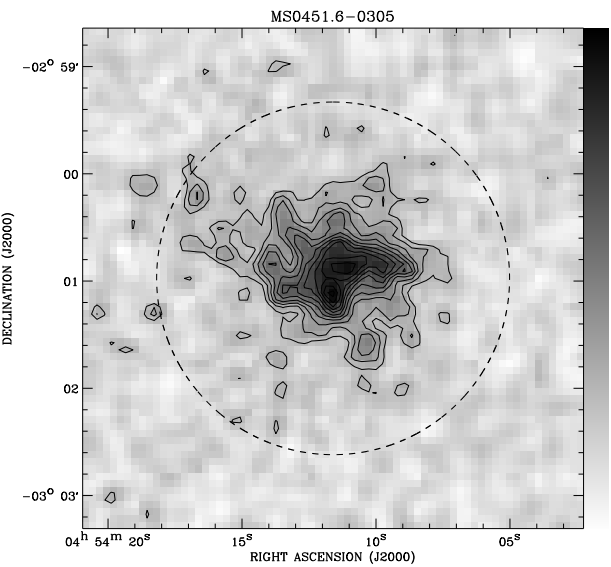
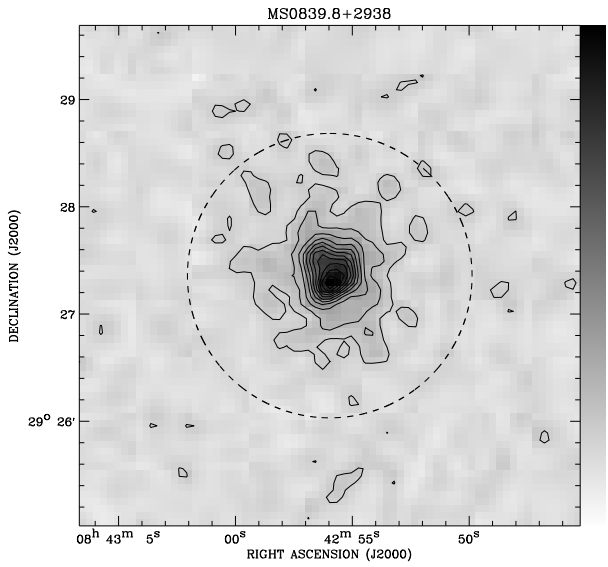
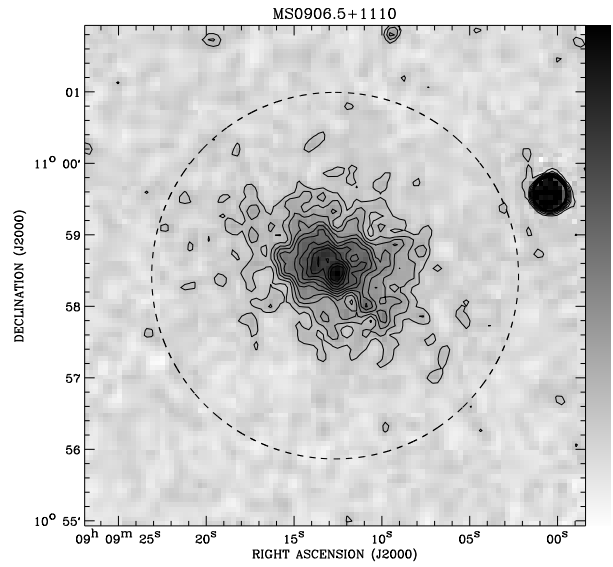
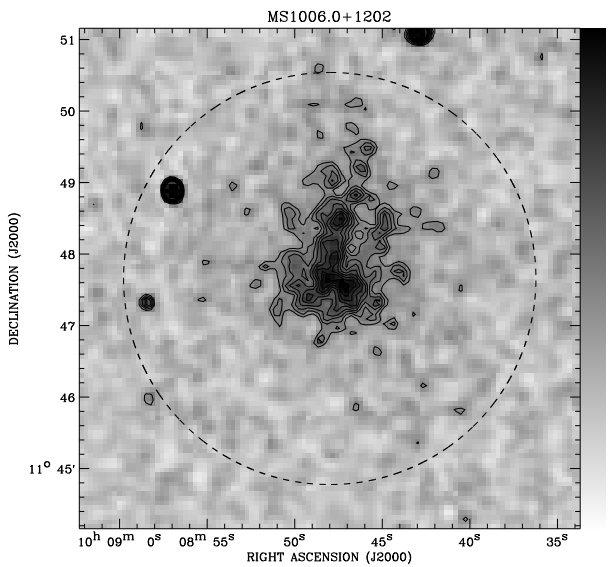
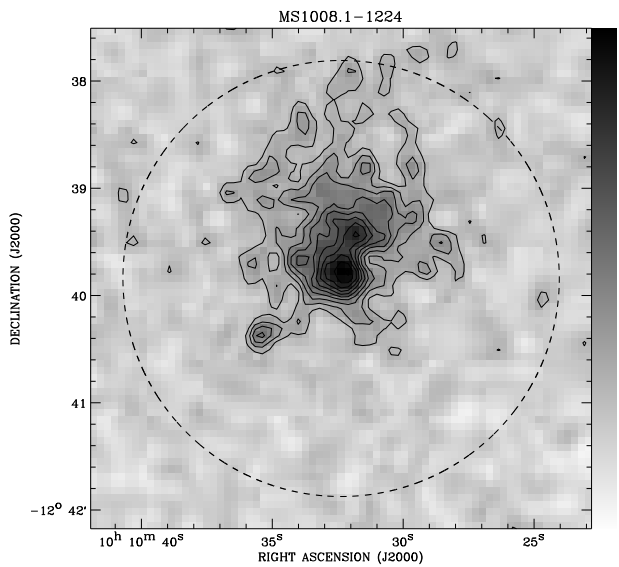
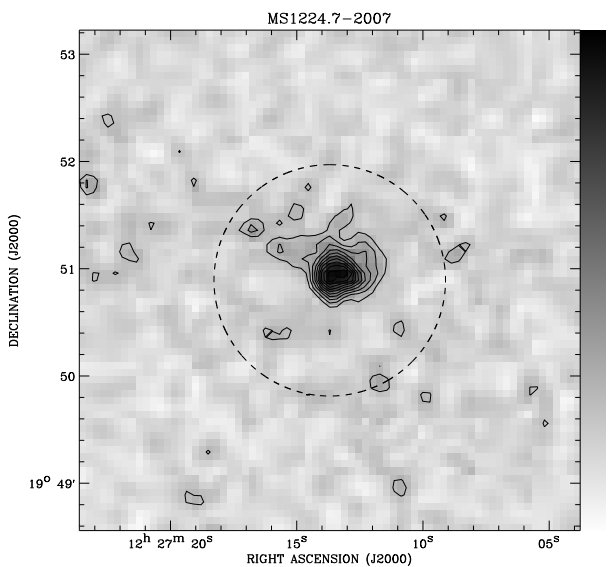
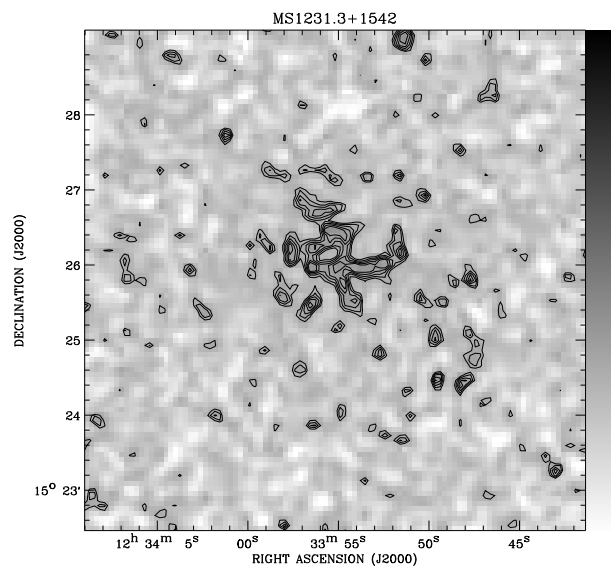
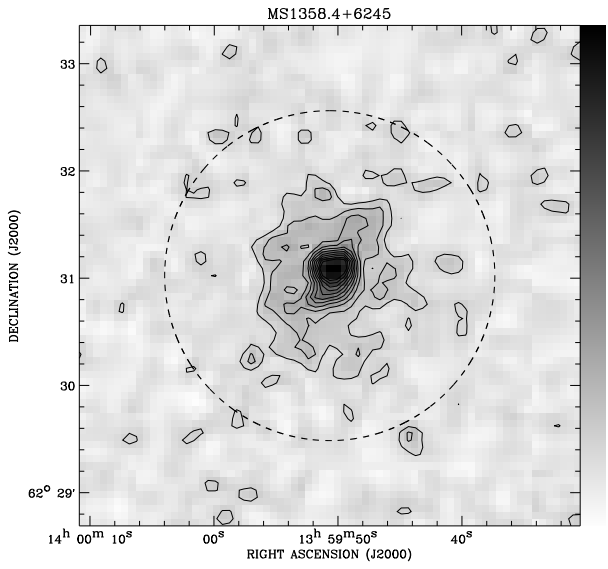
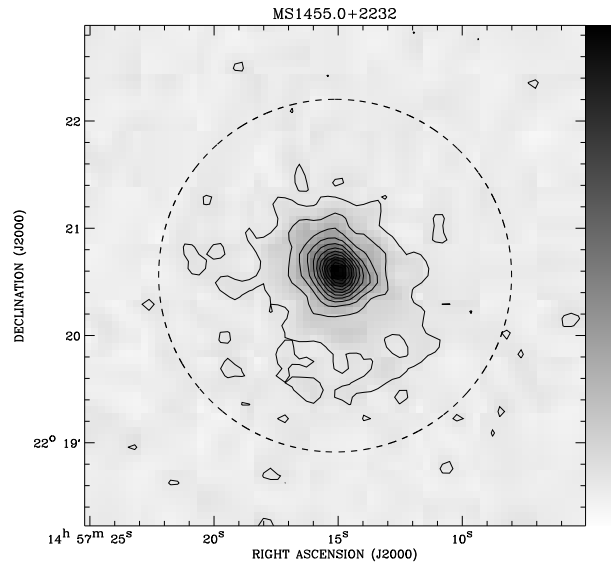
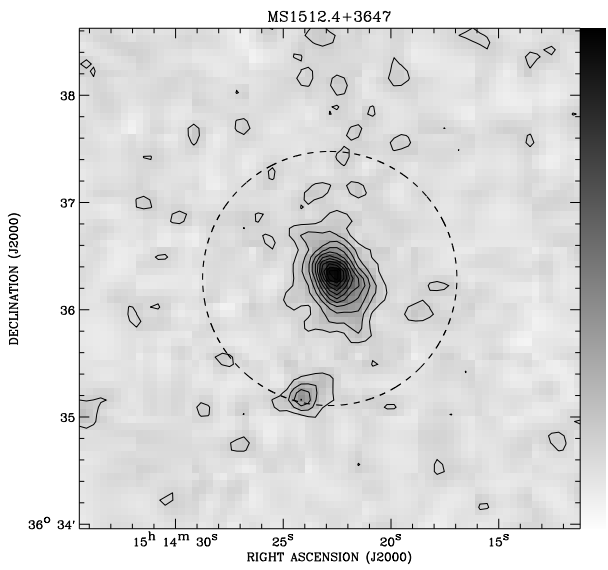
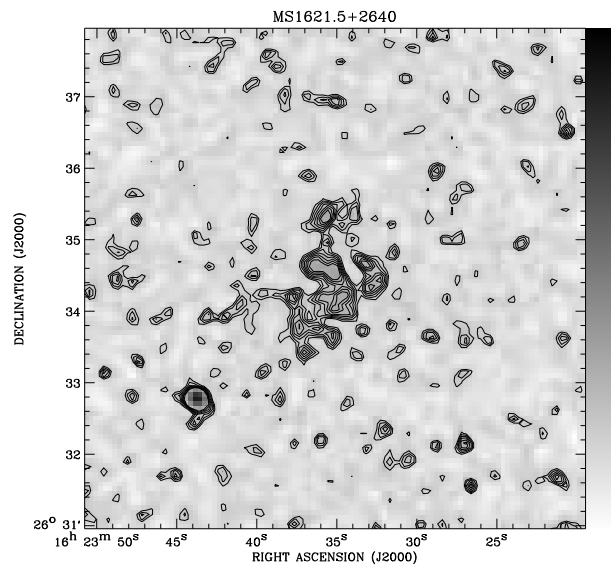
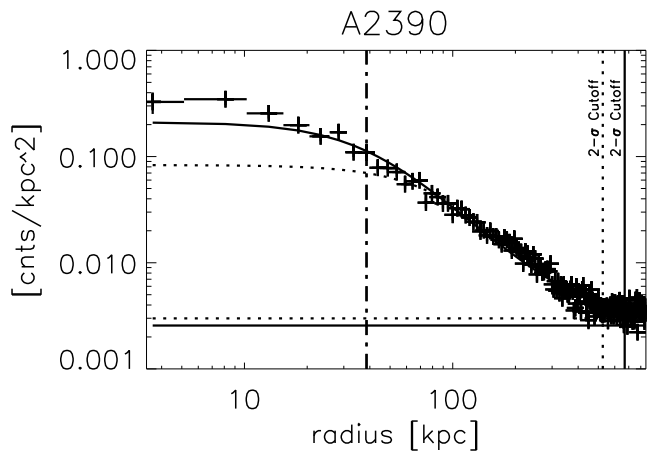
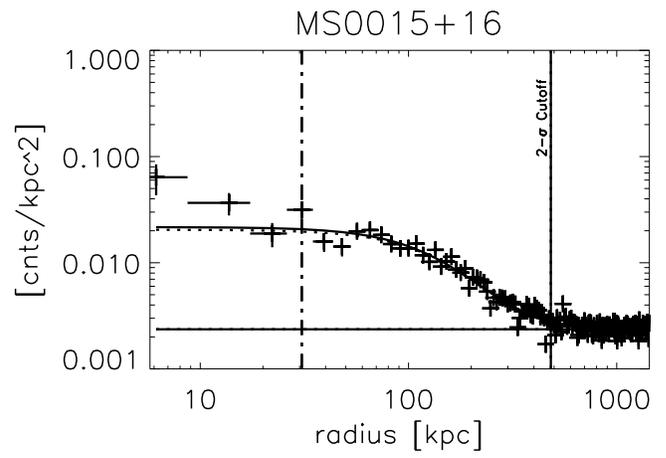
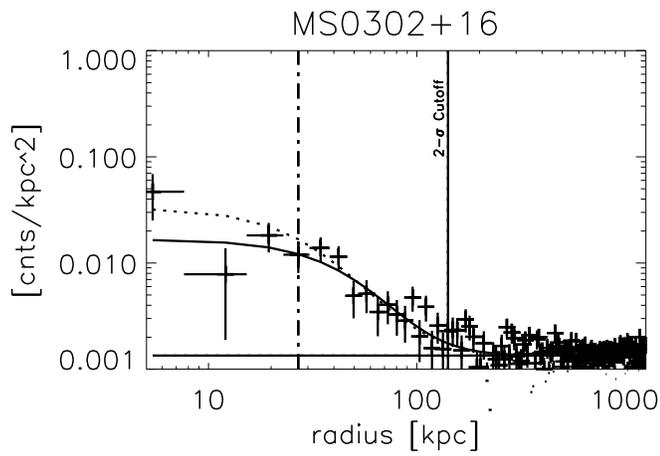
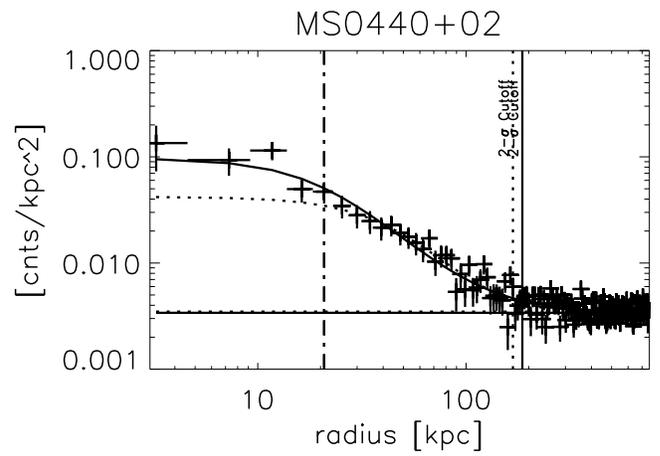
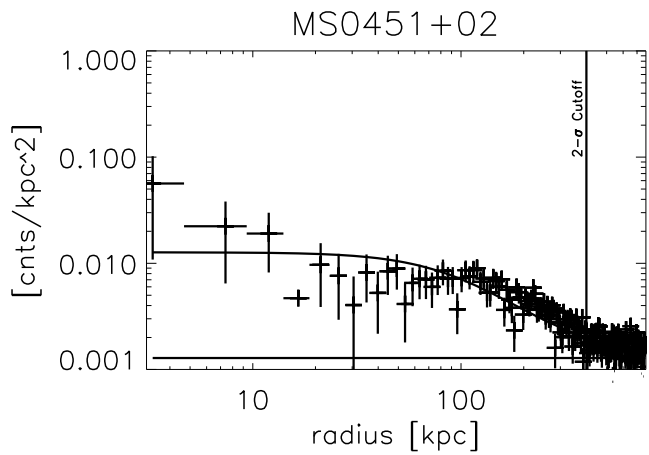
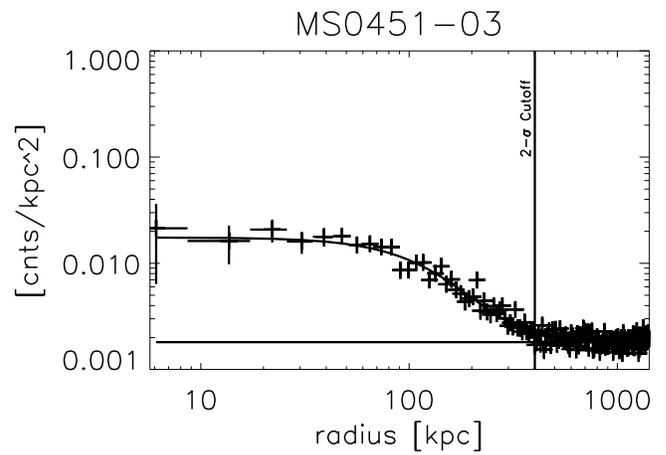
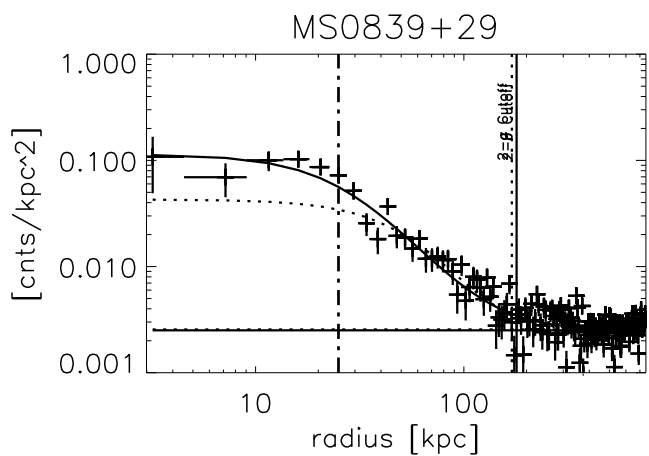
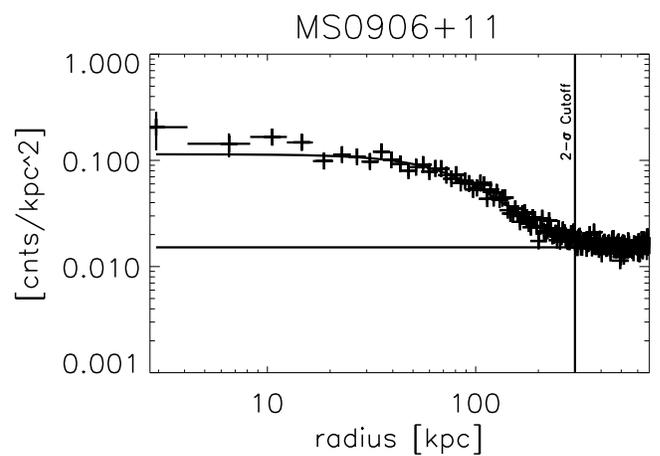
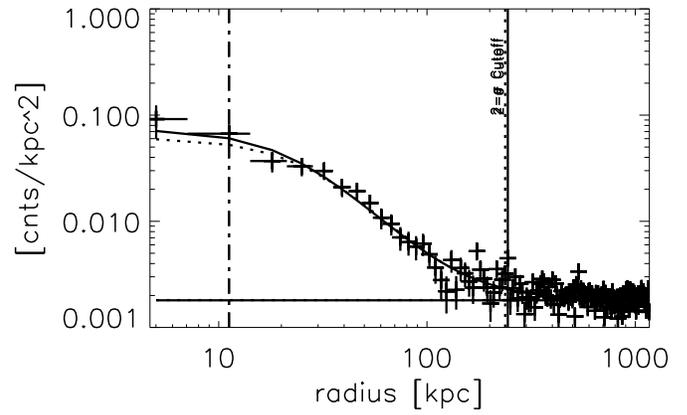
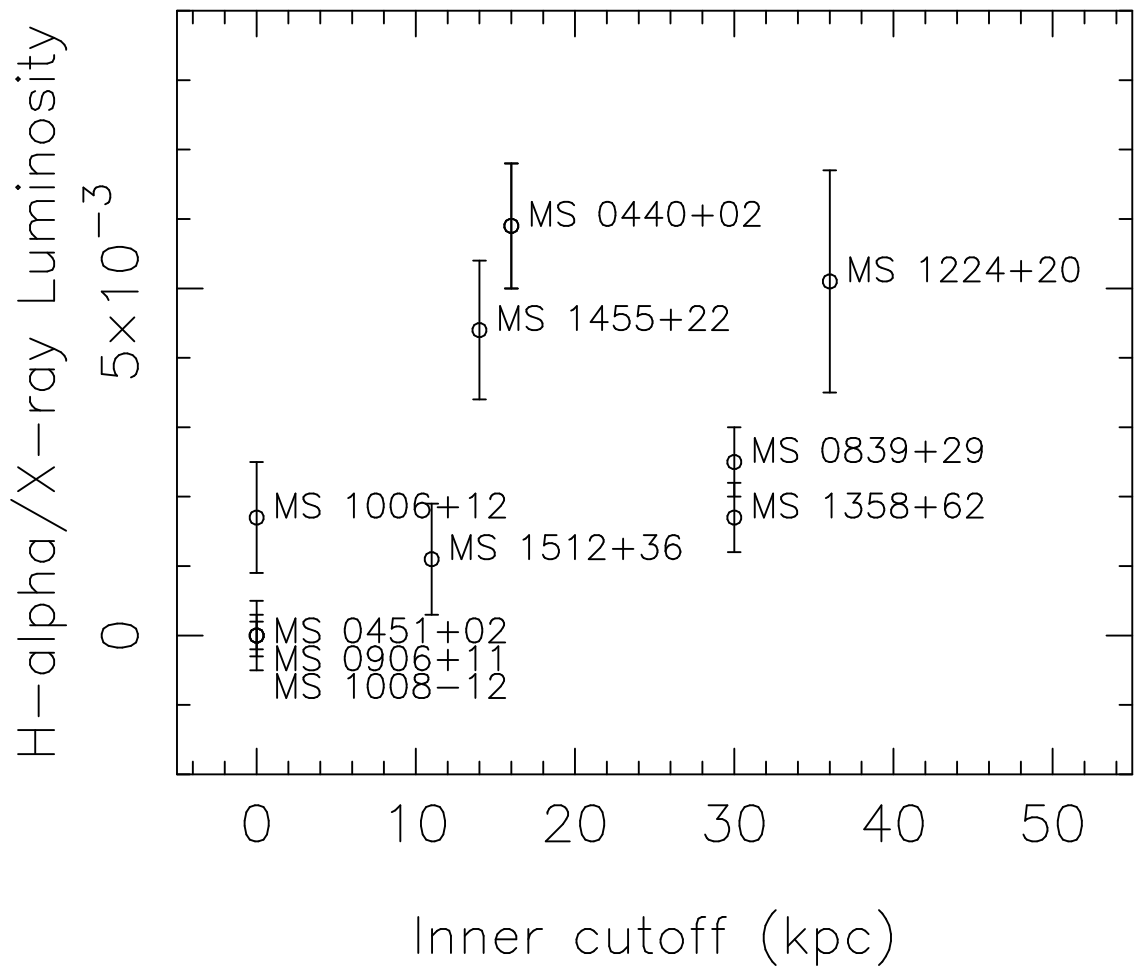
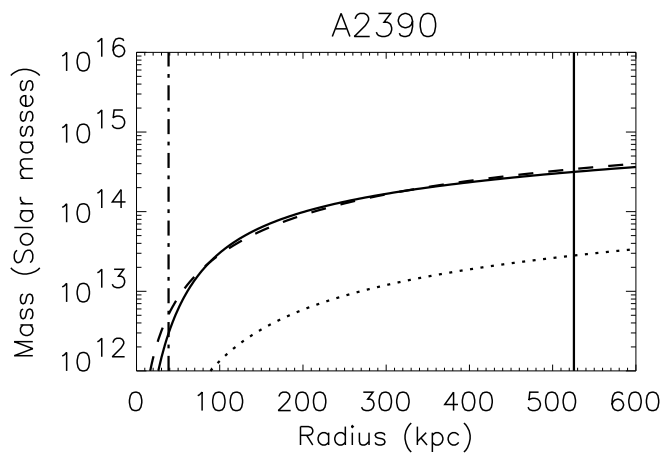
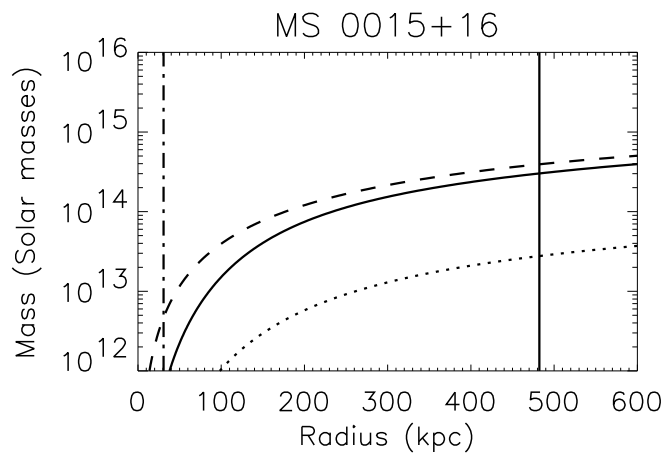
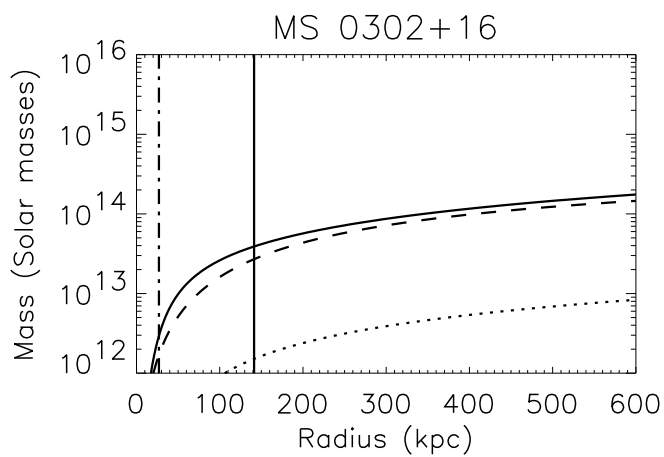
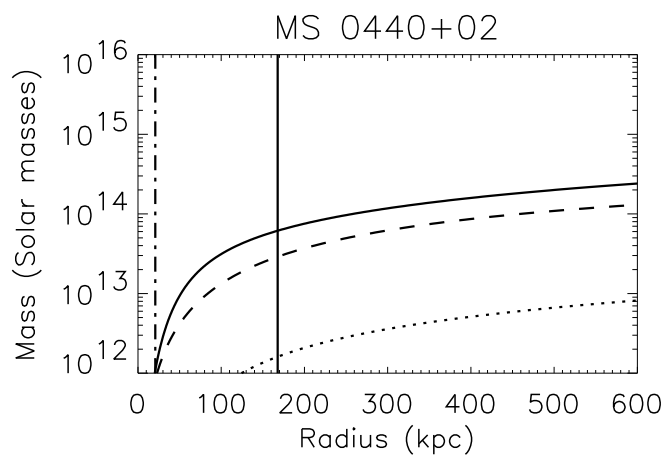
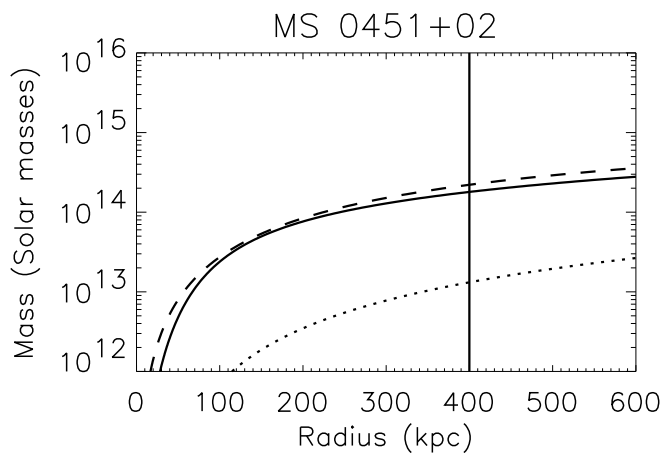
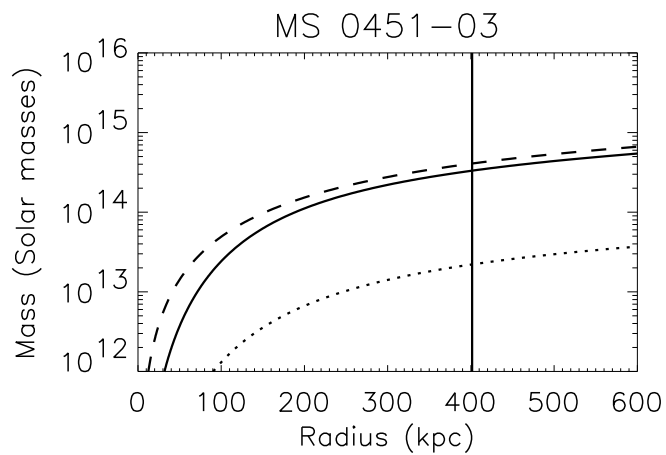
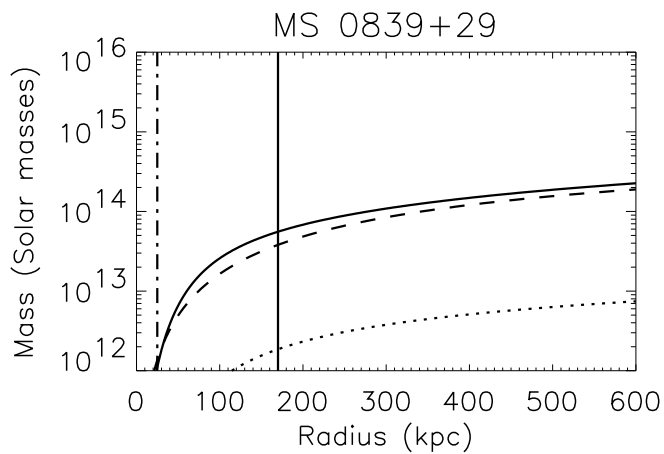
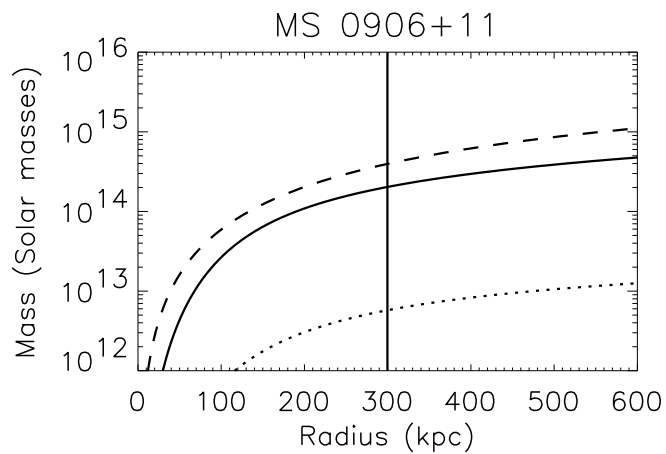
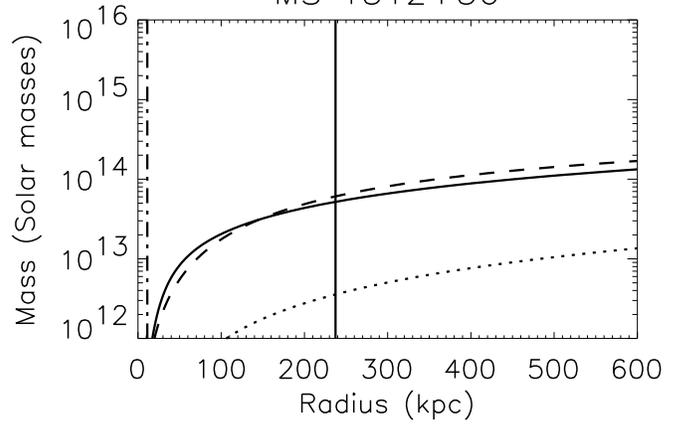
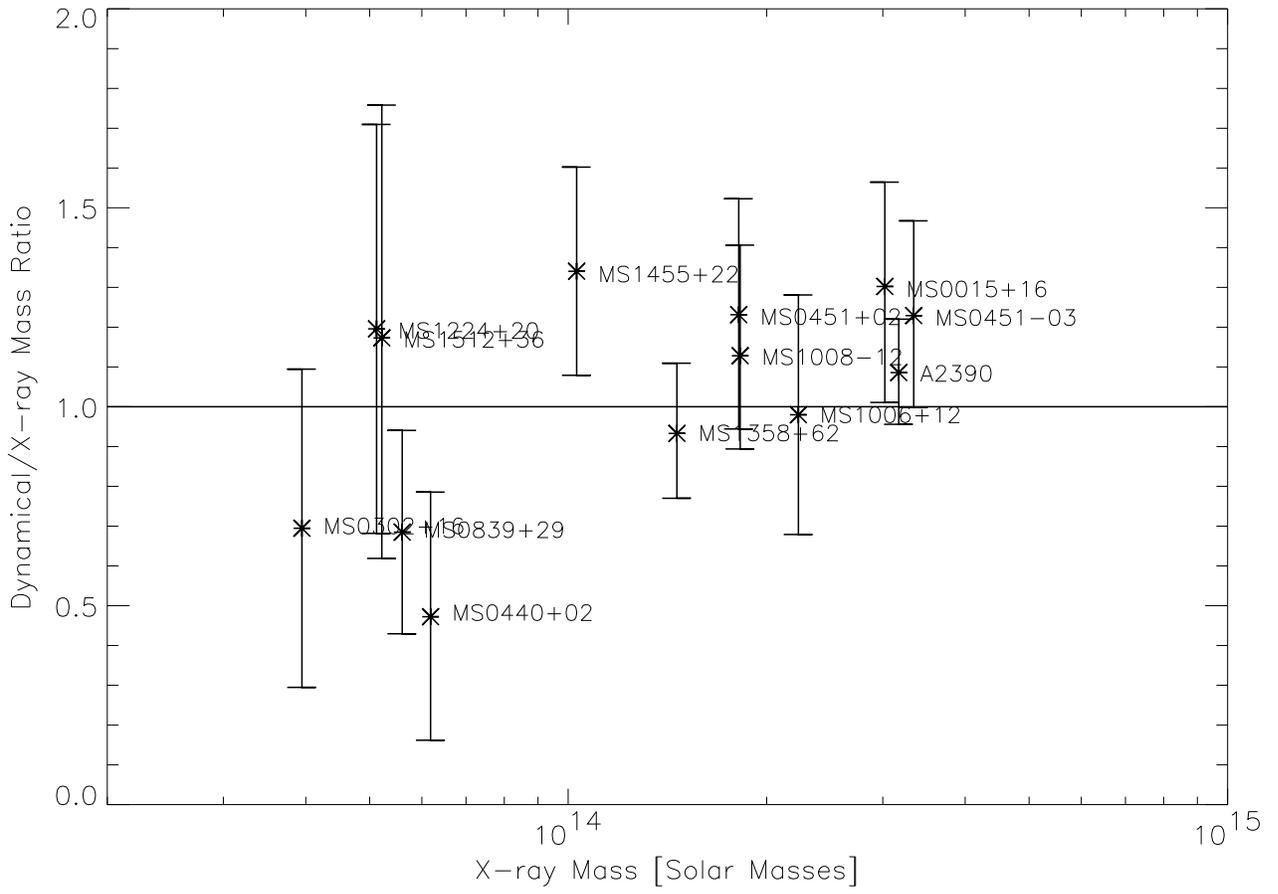
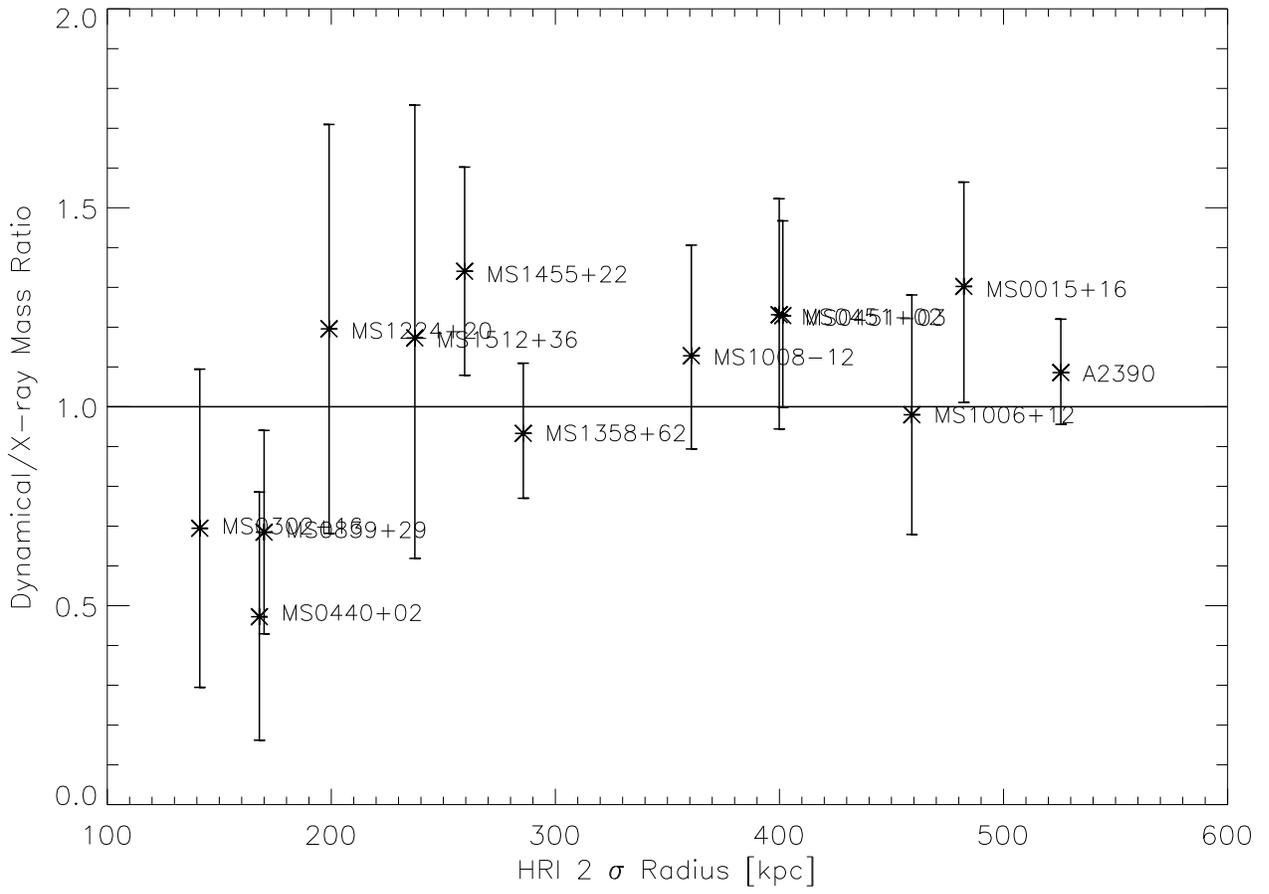
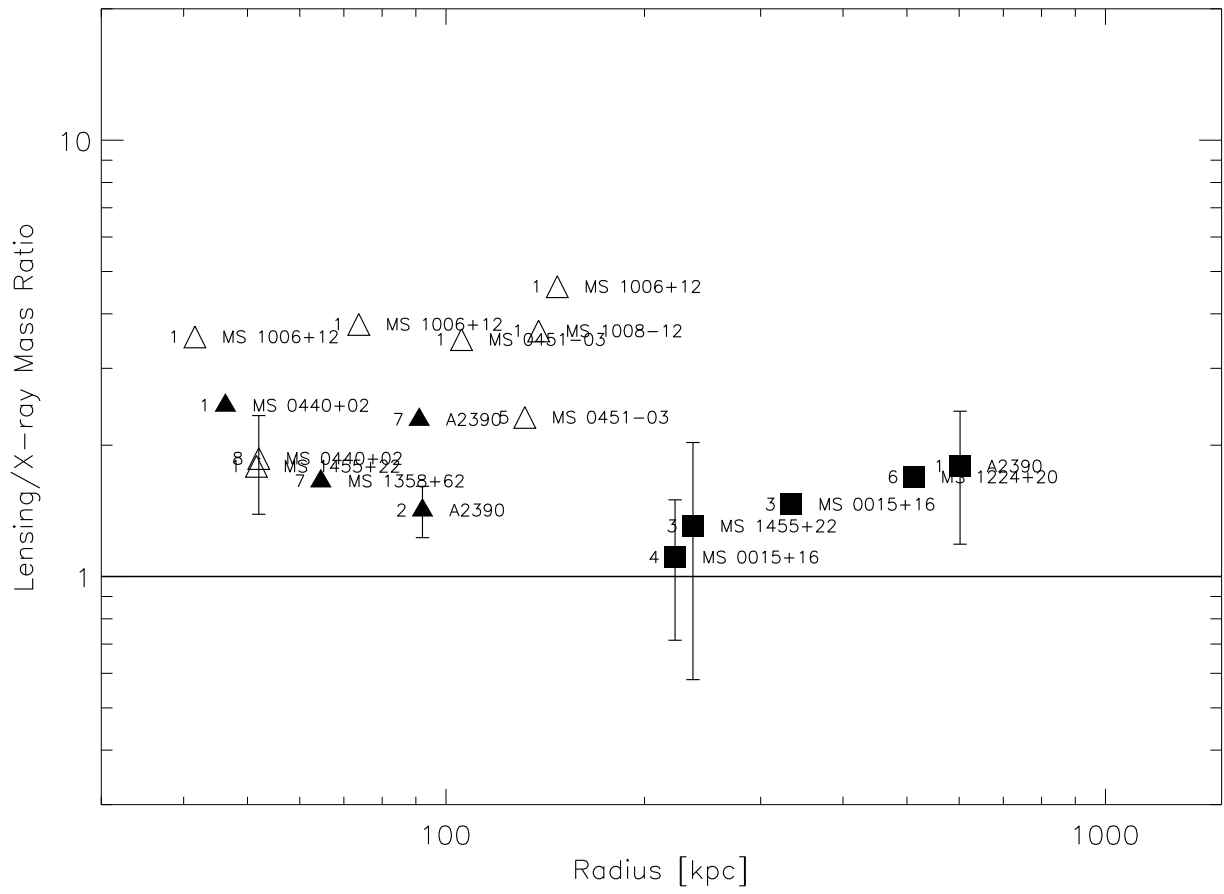
\end{document}